\documentclass[amsmath,amssymb,aps,prb,twocolumn,floatfix,superscriptaddress]{revtex4-1}

\usepackage{graphicx}
\usepackage{dcolumn}
\usepackage{bm}
\usepackage{subfigure}
\usepackage{epstopdf}
\usepackage{enumerate}

\newcommand{\be}{\begin{equation}}
\newcommand{\ee}{\end{equation}}
\newcommand{\ba}{\begin{array}}
\newcommand{\ea}{\end{array}}
\newcommand{\bea}{\begin{eqnarray}} 
\newcommand{\eea}{\end{eqnarray}}

\def\temp(#1){\langle #1\rangle}

\def\tempp(#1){\langle {#1}|}

\def\temppp(#1){|#1\rangle}

\def\ttt(#1,#2){\left(\!\!\ba{c} {#1}\\{#2}\ea\!\!\right)}
\def\spin#1{\ttt#1}

\def\tttt(#1,#2){\left(\!\!\ba{cc} {#1} & {#2}\ea\!\!\right)}

\pdfpageattr {/Group << /S /Transparency /I true /CS /DeviceRGB>>}

\DeclareGraphicsExtensions{.pdf,.png}

\begin{document}

\title{Shot noise in a harmonically driven ballistic graphene transistor}

\author{Y. Korniyenko}
\affiliation{Department of Microtechnology and Nanoscience - MC2,
Chalmers University of Technology, SE-412 96 G\"oteborg, Sweden}

\author{O. Shevtsov}
\affiliation{Department of Physics \& Astronomy, Northwestern University,
Evanston, Illinois 60208, USA}

\author{T. L\"ofwander}
\affiliation{Department of Microtechnology and Nanoscience - MC2,
Chalmers University of Technology, SE-412 96 G\"oteborg, Sweden}

\date{\today}

\begin{abstract}

We study time-dependent electron transport and quantum noise in a ballistic graphene field effect transistor driven by an ac gate potential.
The non-linear response to the ac signal is computed through Floquet theory for scattering states and Landauer-B\"uttiker theory
for charge current and its fluctuations.
Photon-assisted excitation of a quasibound state in the top-gate barrier leads to resonances in transmission
that strongly influence the noise properties.
For strong doping of graphene under source and drain contacts, when electrons are transmitted through the channel via evanescent waves,
the resonance leads to a substantial suppression of noise.
The Fano factor is then reduced well below the pseudo-diffusive value, $F<1/3$, also for strong ac drive.
The good signal-to-noise ratio (small Fano factor) on resonance suggests that the device is a good candidate
for high-frequency (THz) radiation detection.
We show analytically that Klein tunneling (total suppression of back-reflection) persists for perpendicular incidence
also when the barrier is driven harmonically.
Although the transmission is inelastic and distributed among sideband energies, a sum rule leads to total suppression of shot noise.

\end{abstract}

\maketitle

\section{Introduction}

The electronic properties of graphene have attracted considerable attention from the research community
ever since the first experiments with graphene flakes in 2004 \cite{CastroNeto:2009cl,Peres:2010tn,Novoselov:2012hw,2015Nanos...7.4598F}.
The signature characteristics, massless Dirac charge carriers close to the charge neutrality point,
allowed the realisation of a number of interesting physical effects,
for instance Klein tunnelling \cite{Katsnelson:2006kd},
Veselago lensing \cite{Cheianov:2007in},
and the anomalous quantum Hall effect \cite{Novoselov:2005es,Zhang:2005gp,Goerbig:2011un}.
Due to graphene's extreme thinness of just one atom,
its properties can be easily modified by proximity to other materials and it also allows for a tunable charge density.
Fabrication techniques combating defect scattering have been steadily improving over the years,
currently allowing graphene encapsulated in hexagonal boron nitride 
to show ballistic behavior in devices longer
than 1 $\mu$m \cite{Rickhaus:2015cp,Chen:2016ep,2015Sci...348..672Z,Bandurin:2016cp,Crossno:2016iy,Ghahari:2016df}.
The high mobility, tuneable charge density, combined with Dirac electron physics, has elevated graphene to become a promising material
for high-frequency electronics \cite{Schwierz:2010ix,Palacios:2010dw,Glazov:2014cp,Otsuji:2012hn,Koppens:2014dy}.
Graphene-based devices already include field-effect transistors \cite{2012PNAS..10911588C},
frequency mixers \cite{Habibpour:fz} and doublers \cite{Wang:2009jq},
and detectors \cite{Vicarelli:2012ch,Mittendorff:2013gk,Cai:2014hh,Zak:2014gc}.

Possible high-frequency applications have driven a broad theoretical research effort into time-dependent transport with topics covering e.g.
quantum pumping\cite{Prada:2009jo,FoaTorres:2011kf,SanJose:2011fe,Zhu:2011cr,SanJose:2012bw},
electromagnetic response \cite{Mikhailov:2007ft,Mikhailov:2008ig,Syzranov:2008dg,Calvo:2012bg,AlNaib:2014jj,Sinha:2012fx}
and photon-assisted tunnelling \cite{Trauzettel:2007kq,Zeb:2008ka,Rocha:2010ho,Savelev:2012dg,Lu:2012it,Szabo:2013bn,Zhu:2015bd}.
In high-frequency devices, time-dependent electric field of frequency $\Omega$ induces sidebands in energy space
separated by multiples of energy quantum $\hbar\Omega$.
Interference of quasiparticle scattering paths between the sidebands
is therefore important \cite{Lu:2012it,Szabo:2013bn,Zhu:2015bd,Korniyenko:2016ct,Korniyenko:2016hg}.
In our previous papers \cite{Korniyenko:2016ct,Korniyenko:2016hg} we examined in detail
the linear conductance of a ballistic graphene transistor with an ac-driven top gate.
Scattering via quasibound states under the gate induces resonances in selected sideband amplitudes,
and we have identified two resonant scattering mechanisms:
double barrier tunnelling (between contacts and top barrier) for high doping of the contacts and Breit-Wigner/Fano resonances for low doping.
We showed  that based on these resonances the device can be used as a detector in terahertz (THz) frequency range for weak driving of the gate
or as a frequency multiplier for strong driving.
In this paper we develop further our model based on Floquet theory and Landauer-B\"uttiker scattering
formalism \cite{BAGWELL:1992wx,1998PhRvB..5812993P,Platero:2004ep,2005PhR...406..379K} (adequately generalized for graphene)
to include shot noise.

Ballistic Klein tunnelling is associated with low noise desirable in electronic devices.
Since the effect remains robust even for high doping of contacts, transport at the charge neutrality point
is characterised by a mixture of evanescent waves and Klein tunnelling.
It leads to a universal minimal conductivity\cite{Katsnelson:2006gl} of $4e^2/\pi h$
and the Fano factor\cite{Tworzydio:2006hw} reaches a local maximum of $1/3$.
This sub-Poissonian value for noise coincides with that of disordered diffusive metals and has been verified experimentally \cite{Danneau:2008kg}.
Shot noise, being a measure of current-current correlations, potentially contains more information than can be extracted from the dc conductance.
It has therefore attracted considerable attention in the study of electronic quantum transport, see for instance the review in Ref.~\onlinecite{2000PhR...336....1B}.
In the context of graphene, photon-assisted shot noise was recently measured in the diffusive electron transport regime \cite{Parmentier:2016ec}.
It was shown that shot noise signatures of radiation in this system could be extended to the higher THz frequency range.
In another recent experiment \cite{Laitinen:2016ht}, shot noise was utilized to extract detailed information about contact doping
and the doping profile across suspended graphene field effect transistors.
Several theoretical studies of shot noise in graphene have also appeared recently.
Signatures of Fabry-P\'erot interferences in the shot noise have been investigated,
both the zero frequency noise \cite{Rocha:2010ho} and the finite frequency noise \cite{Hammer:2013gb}.
In the latter case, it was shown that the noise power oscillates with frequency on a scale set by the Fabry-P\'erot energy scale $L/\hbar v_F$,
where $v_F$ is the Fermi velocity and $L$ is the distance between source and drain contacts.
Noise was also calculated for adiabatic \cite{Zhu:2011cr} or non-adiabatic quantum pumps in graphene \cite{Zhu:2015bd}.
The current work complements these works and focuses on the signatures in shot noise of the different resonant scattering mechanisms
identified in Refs.~\onlinecite{Korniyenko:2016ct,Korniyenko:2016hg}, and in particular the usefulness of these resonances in high-frequency (THz) radiation detection
in a set-up sketched in Fig.~\ref{tr}.

\section{Model}\label{sec:model}

\begin{figure}[t]
\includegraphics[width=0.9\columnwidth]{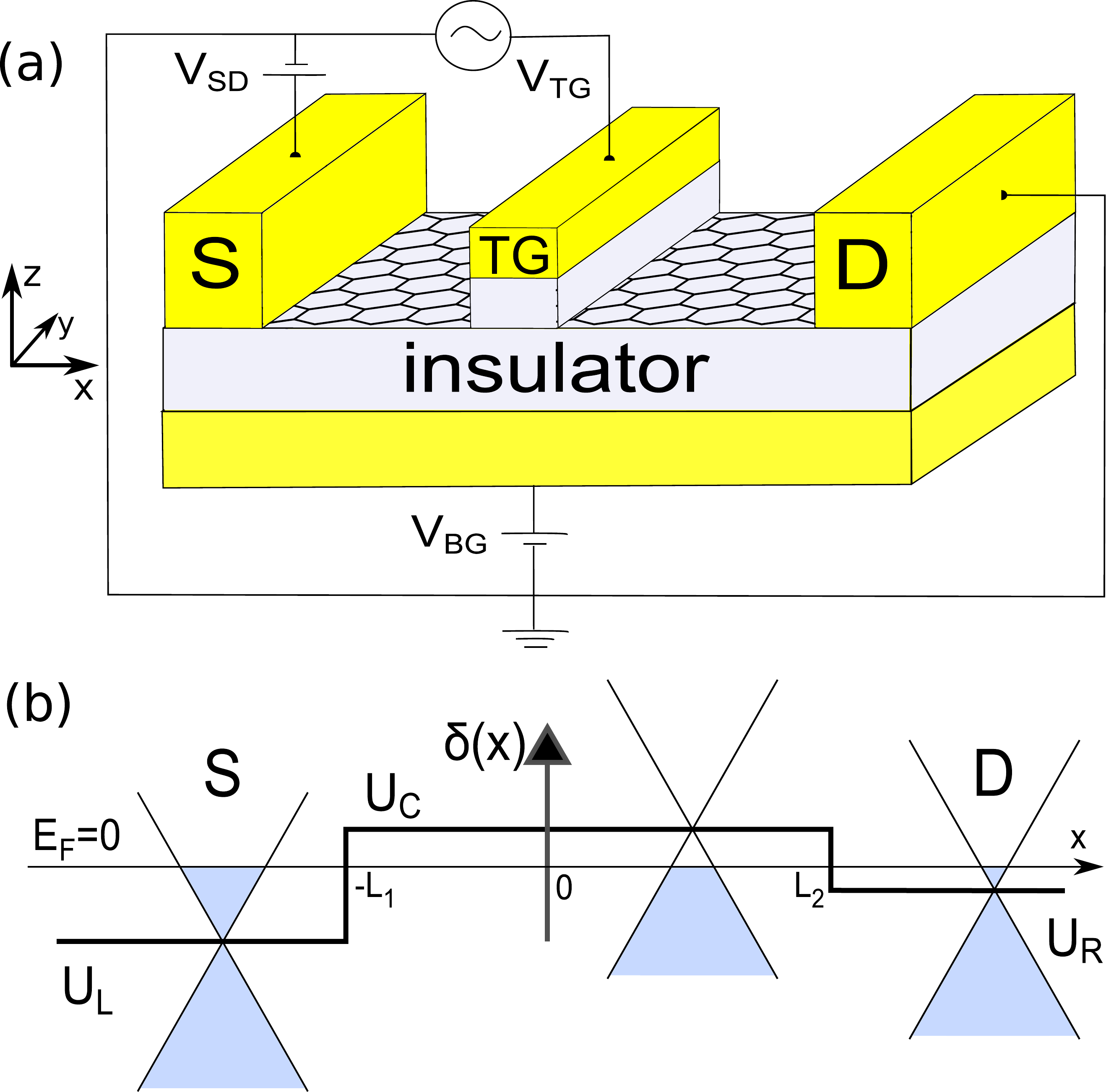}
\caption{(a) Schematics of a graphene field effect transistor, where a back gate (BG) controls doping of the channel,
a small source (S) - drain (D) bias is applied to generate the current, which is controlled by the top gate (TG) dc and ac signals.
(b) Potential landscape, including doping of the leads by the source and drain metallic electrodes.}
\label{tr}
\end{figure}

Our aim is to investigate intrinsic noise properties of the ballistic device depicted in Fig.~\ref{tr}(a),
excluding the extrinsic effects of surrounding circuitry eventually present in an experiment.
To this end we utilize a minimal model of the device based on Floquet theory for Dirac quasiparticle scattering states
combined with a Landauer-B\"uttiker theory for transport.
We shall focus on the shot noise at zero temperature in parameter regimes corresponding to the resonances
discussed in detail in Ref.~\onlinecite{Korniyenko:2016ct,Korniyenko:2016hg}.
The device depicted in Fig.~\ref{tr}(a) consists of a graphene sheet contacted by source and drain electrodes.
A harmonic signal is assumed to be applied to the top-gate and we compute the complete non-linear response
to this signal within our minimal model. The observable we focus on is the current and its fluctuations (noise)
between source and drain in linear response to a voltage $V_{\mathrm{SD}}=V_{\mathrm{S}}-V_{\mathrm{D}}$. Note that in the following
we use $V_{\mathrm{S}}$, since the drain is grounded in Fig.~\ref{tr}(a).

The device is assumed to have an ideal contact geometry, invariant in the transverse $y$-direction,
resulting in a quasi-one-dimensional potential landscape between source and drain ($x$-direction),
\bea
U(x) &=& U_L\theta(-L_1-x) + U_R\theta(x-L_2) \nonumber\\
&&+\,U_C\left[ \theta(x+L_1)-\theta(x-L_2)\right],
\label{doping_profile}
\eea
as sketched in Fig.~\ref{tr}(b). The function $\theta(x)$ is the Heaviside step function.
The shifts $U_L$ and $U_R$ take into account doping of graphene by deposited metallic electrodes and do not change
under back-gate potential sweep and small source-drain bias $V_{\mathrm{S}}$ \cite{Huard:2008hx}.
The potential in the channel region (between source and drain), on the other hand, is controlled by a back gate setting the Dirac point energy $E_D=U_C$.
We measure energy from the Fermi level of metallic contacts $E_F=0$.
Thus, for $U_C=0$ the Dirac point in the graphene channel is aligned with the Fermi energy of the leads.
Below we shall consider a symmetric set-up with $U_L=U_R=U$ and $L_1=L_2=L/2$.

We assume that the contacts are smooth on the atomic scale, but sharp on the wavelength associated with the energy of Dirac electrons in the channel $\lambda_D=\hbar v_F/|E-U_C|$, where $v_F$ is the Fermi velocity.
In this approximation, the potential changes across the device in a step-like fashion on the scale of $\lambda_D$, as in Eq.~(\ref{doping_profile}).
This approximation also allows us to disregard intervalley scattering and include only one Dirac point in our Hamiltonian. For calculation of current and noise we include a factor of four for spin and valley degeneracy. In addition, we assume that the top gate is narrow on the scale of $\lambda_D$, allowing us to treat it as a delta potential in our model, see also Fig. 1(b).
The effective low-energy Hamiltonian then has the form
\be
\mathcal{H}=-i\sigma_x\nabla_x+\sigma_y k_y+\left[Z_0+Z_1\cos(\Omega t)\right]\delta(x)+U(x),
\label{Hamiltonian}
\ee
where we have set the Fermi velocity in graphene equal to unity, $v_F=1$, and $\hbar=1$. Note that the energy scale in these units is set by $1/L$.
Here $Z_0$ is the strength of the static part of the delta barrier, while $Z_1$ of its dynamic part.
Pauli matrices in pseudospin space (A-B sublattice degree of freedom) are denoted $\sigma_{x}$ and $\sigma_y$.
We assume the device to be wide in the transverse direction to disregard any finite-size effects along $y$ axis.
Together with translational invariance it allows the transverse momentum $k_y$ to be conserved during scattering.
Below we will often express $k_y$ in terms of an impact angle $\varphi$ via the relation $k_y=|U|\sin\varphi$.

Given the above Hamiltonian, we need to solve the time-dependent Dirac equation
\begin{equation}
\mathcal{H}\psi(x,k_y,t)=i\partial_t\psi(x,k_y,t).
\label{Dirac_eq}
\end{equation}
We utilize the periodicity of the Hamiltonian in the time domain and use a Fourier decomposition to build the Floquet ansatz
\begin{equation}
\psi(x,k_y,t) = e^{-iEt} \sum\limits_{n=-\infty}^{\infty} \psi_n(x,k_y,E) e^{-in\Omega t}.
\end{equation}
The quasienergy $E$ is set by the energy of the incoming electron from the lead.
As the charge carrier exchanges energy quanta $n\Omega$ ($n$ integer) in the top-gate barrier, the wavefunction acquires amplitudes
at the sideband energies $E_n=E+n\Omega$.
The Dirac equation is then rewritten as a matrix differential equation in sideband space.
Solutions to it are obtained through wavefunction matching at interfaces between regions with different potentials \cite{Korniyenko:2016ct,Korniyenko:2016hg}.
The solutions can be collected into a Floquet scattering matrix $S_{\alpha\beta}(E_n,E_m)$ for scattering from contact $\beta$ at energy $E_m$
to contact $\alpha$ at energy $E_n$, where in our case the contact indices $\alpha,\beta\in\{\mathrm{S,D}\}$, see Fig.~\ref{tr}.

Adapting the Landauer-B\"uttiker scattering formalism,
we can express the expectation values for charge current and noise in the system in terms of the Floquet scattering matrices.
The key steps are outlined in Appendix~\ref{App_A}.
In the following we present an analysis of differential noise $\mathbb{N}$, computed with Eq.~(\ref{eq:Nfinal}),
and compare it to the dc linear conductance $G_0$, computed with Eq.~(7) in Ref.~\onlinecite{Korniyenko:2016hg}.
To this end we mainly focus on their ratio, the Fano factor, defined in Eq.~(\ref{eq:F}), which measures the deviation from Poissonian noise.
We assume that temperature is the lowest energy scale (we set $T=0$),
while other parameters are intentionally chosen as in Ref.~\onlinecite{Korniyenko:2016hg}, where these choices were thoroughly motivated
both by experimental relevance and subdivision into most interesting transport regimes.

\subsection{Dc characteristics}

\begin{figure}[t]
\includegraphics[width=\columnwidth]{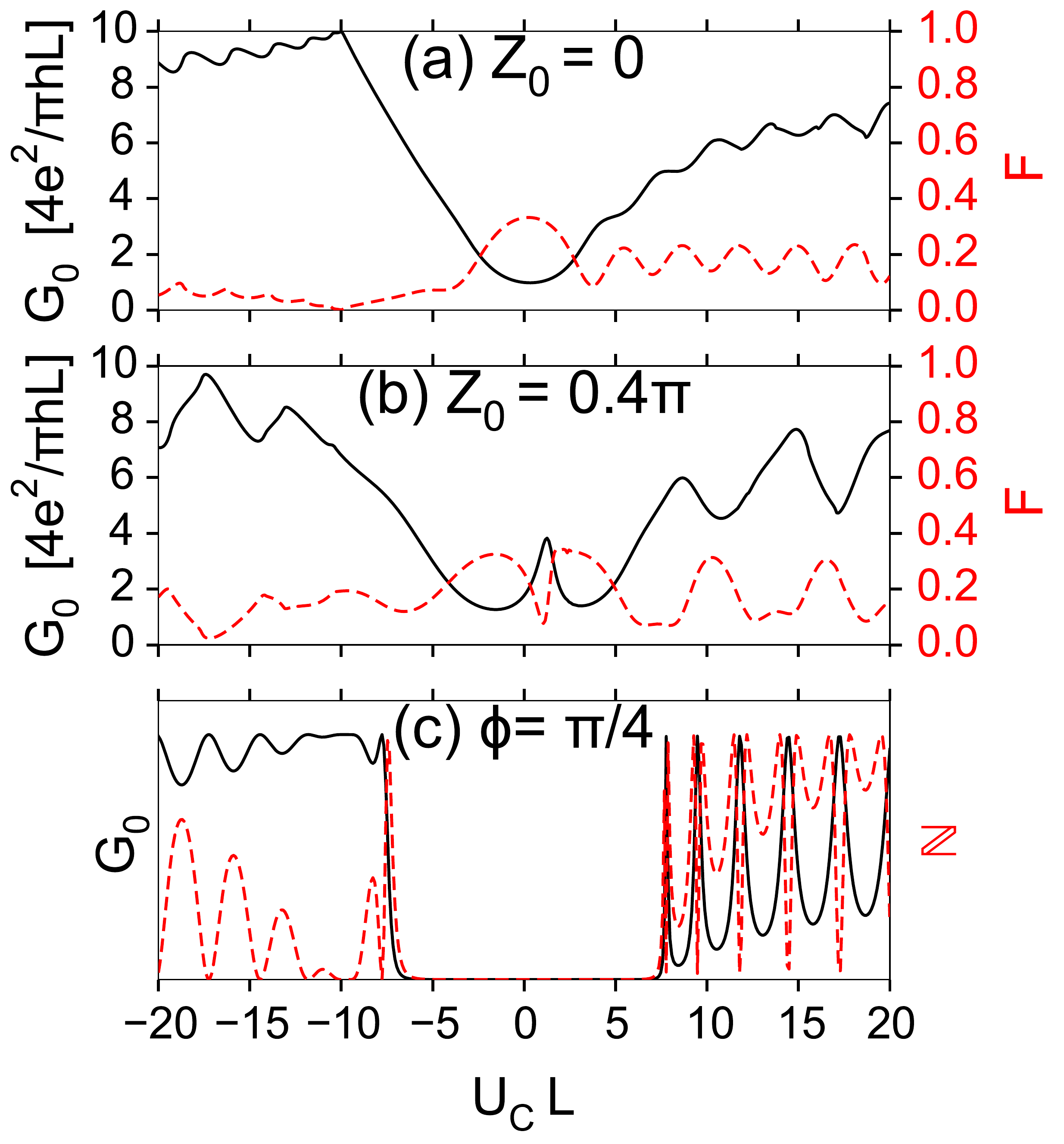}
\caption{Dc conductance (solid) and dc Fano factor (dashed) at finite doping $U=-10/L$ with the gate (a) off $Z_0=0$ and (b) on at $Z_0=0.4\pi$.
(c) The same as in (a) but for one particular impact angle $\varphi=\pi/4$.
In (c) the conductance and noise are rescaled (arb. units) to show peak-dip correspondence.}
\label{fig:dc}
\end{figure}

In the absence of external ac drive, transport is elastic. The zero temperature shot noise expression is then simplified to a well-known result
\be
\mathbb{N} = \left.\frac{e^3}{2\pi h}\int\limits_{-\infty}^{\infty}dk_y\, T(k_y,E)\left[1-T(k_y,E)\right]\right|_{E=E_F},
\ee
and the differential Fano factor is given by
\be
F=\frac{\mathbb{N}}{eG_0} = \left.\frac{\int\limits_{-\infty}^{\infty}dk_y\, T(k_y,E)\left[1-T(k_y,E)\right]}{\int\limits_{-\infty}^{\infty}dk_y\, T(k_y,E)}\right|_{E=E_F}.
\ee
Let us first analyse the noise with no dc gate applied ($Z_0=0$).
For strong doping of leads $U\gg U_C$, the transport channel is characterised by evanescent waves.
When transport is exactly at the Dirac point (for $U_C=0$), the conductance (computed per unit width of the device in the $y$-direction)
has a minimum $G_0=\frac{4e^2}{\pi h L}$,
and the Fano factor approaches its maximum value of $F=1/3$, see Fig.~\ref{fig:dc}(a).
This sub-poissonian value coincides with the diffusive metal result\cite{1992PhRvB..46.1889B}
and is called the pseudo-diffusive transport regime \cite{Tworzydio:2006hw}.
For higher doping of the channel $|U_C|>0$, it becomes more transparent and noise correlations are suppressed further.
There exists an extreme point at $U_C=U$ where $T=1$ for all angles of incidence (all $k_y$) and the noise is suppressed to zero.
Another notable feature is the oscillations of both the conductance and the Fano factor on the scale of $1/L$.
These oscillations are associated with Fabry-P\'erot resonances\cite{Hammer:2013gb} induced by two partly reflecting mirrors at
the interfaces between the channel and the contacts [at $x=-L_1$ and $x=L_2$ in Fig.~\ref{tr}(b)].
We note that for fixed $k_y$, local maxima in conductance have corresponding minima in the Fano factor and vice versa.
After integration over $k_y$, most maxima and minima still coincide, see Fig.~\ref{fig:dc}(a).
We can quantify the phenomenon analytically.
Denoting $T'=\partial T/\partial U_C$ we can analyze the behavior of noise at conductance extrema $G'=0$ by looking at first and second derivatives of the observables.
Disregarding the integration over transverse momentum $k_y$ we find
\begin{align}
\mathbb{N}'&\propto[T(1-T)]'=0=G',\\
\mathbb{N}''&\propto [T(1-T)]''=T''(1-2T)\propto G''(1-2T).
\end{align}
The first equation shows that the extrema positions of conductance and noise coincide.
At high channel transparencies ($T>0.5$) the curvature of noise has opposite sign compared to that of conductance at its extrema.
For $U_C$ not too close to zero, evanescent modes (low transparency channels) do not contribute much to conductance,
while Klein tunnelling ensures high transparency for open channels.
Extrema positions then correspond well between conductance and noise for the entire range of doping $U_C$.

When the top gate dc potential is non-zero, $Z_0\ne 0$, the charge carriers can scatter resonantly through a quasibound state in
the delta barrier with energy\cite{Korniyenko:2016ct}
\be
\label{eq:bound}
E_b = U_C - \mbox{sgn}(Z_0) |k_y| \cos Z_0.
\ee
For high doping of contact $U\gg U_C$, the resonance happens in an otherwise evanescent wave region.
The transmission coefficient is then resonantly enhanced resulting in a pronounced conductance peak and a corresponding valley in the noise,
c.f. near $U_C=0$ in Fig.~\ref{fig:dc}(b). Thus, on resonance the Fano factor is greatly suppressed.
We note that the transmission enhancement is maximal for a symmetric setup $L_1=L_2$ (which we focus on in this work),
and gets weaker with increasing asymmetry.
The delta barrier introduces a phase shift to the scattered pseudospinor states which results in shifts of the Fabry-P\'erot interference pattern
for large $U_C$, as compared with the case without the barrier.

\section{Results}\label{sec:results}

\begin{figure}[tb!]
\includegraphics[width=\columnwidth]{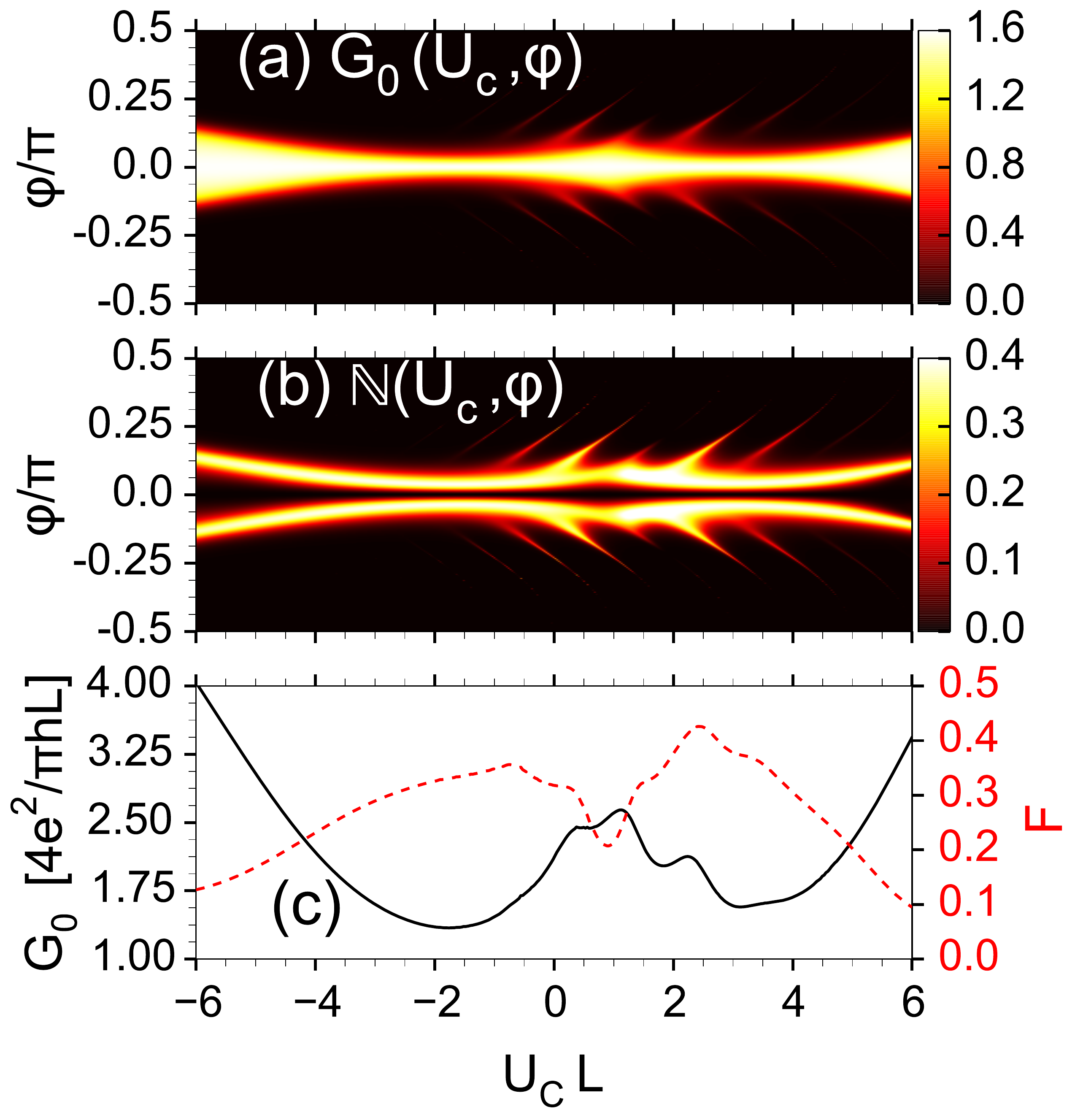}
\caption{(a) Angle-resolved dc conductance $G_0$, (b) angle-resolved differential shot noise $\mathbb{N}$,
and (c) angle-integrated dc conductance and differential shot noise.
The model parameters are $Z_0=0.4\pi$, $Z_1=0.4$, $U=-10/L$, and $\Omega=1/L$.}
\label{fig:double}
\end{figure}

In the previous section we have seen how a tunnelling resonance as a rule of thumb effectively lowers the noise.
Let us first discuss in general terms what is expected when an ac drive on the top gate is applied.
As a result of the harmonic drive, multiple sidebands are generated in energy space.
In effect many additional resonant scattering processes are introduced.
For instance, compared to the static case above, the resonance peak is now split into many peaks,
roughly separated in energy space by $\hbar\Omega$ from each other.
At this point we should note that the resonance combs in transmission to different sidebands coincide, thus producing a single comb in conductance,
see Fig.~\ref{fig:double}(a).
For the noise, it describes current fluctuations between different scattering processes.
Since additional scattering processes are introduced under ac drive, we expect as a rule of thumb that the noise is enhanced as compared with dc.
A careful examination of Eq.~(\ref{eq:Nfinal}) reveals that the overlap of combs in transmission functions also produces a resonant peak comb in the noise,
as shown in Fig.~\ref{fig:double}(b).
Thus, since the main transmission resonance peak near $U_C=0$ is now split into several,
its weight gets redistributed, resulting in a generally higher shot noise.
The Fano factor for high contact doping $U\gg U_C$, displayed in Fig.~\ref{fig:double}(c),
is enhanced in ac compared with dc [c.f. Fig.~\ref{fig:dc}(b)], but stays below the sub-poissonian value of $F=1/3$ around $U_C=0$.

\begin{figure}[tb!]
\includegraphics[width=\columnwidth]{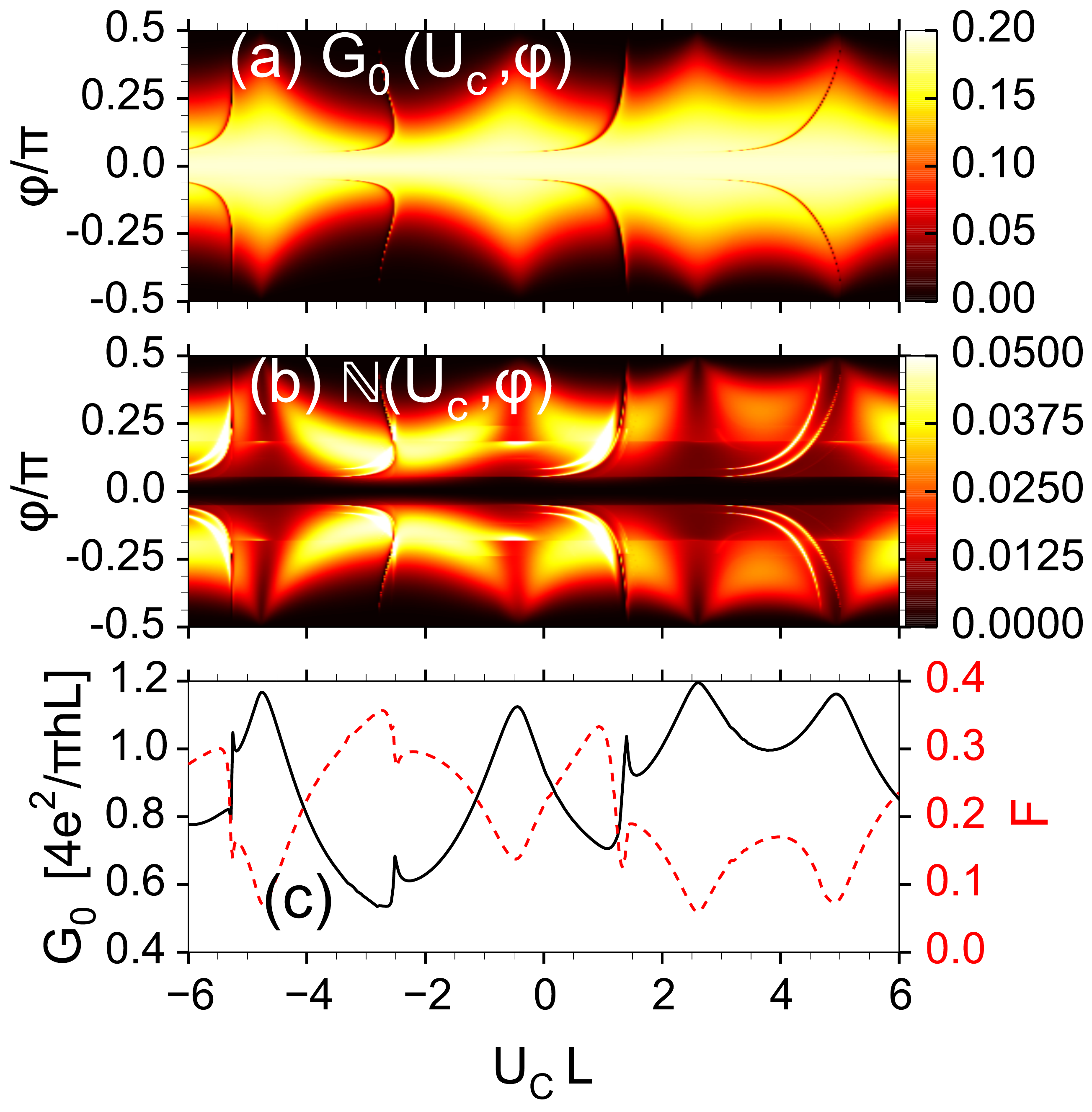}
\caption{(a) Angle-resolved dc conductance $G_0$, (b) angle-resolved differential shot noise $\mathbb{N}$,
and (c) angle-integrated dc conductance and differential shot noise.
The model parameters are $Z_0=0.4\pi$, $Z_1=0.4$, $U=1.2/L$, $\Omega=1/L$. }
\label{fig:fano}
\end{figure}

In the case of low contact doping, pseudospinor mismatch between scattering states is small and the picture is dominated by Klein tunnelling for open channels,
combined with Fano and Breit-Wigner resonance lines induced by the quasibound state \cite{Korniyenko:2016hg}.
In the angle-resolved map of Fig.~\ref{fig:fano}(a), evanescent wave regions manifest themselves as horizontal lines.
Since the critical angle $\phi_c^n$ differs between sidebands $\phi_c^n=\arcsin\left|(n\Omega-U)/U\right|$,
multiple horizontal features are present in the figure which is more evident in noise, see panel (b).
The dc conductance component experiences sharp dip-peak structures due to the Fano and Breit-Wigner resonances.
Noise in the vicinity of corresponding resonances contains extrema (maxima or minima)
due to the fluctuations between several scattering processes affected by these resonances.
The dc conductance only sums over individual sideband transmissions, while noise includes interference terms between different scattering amplitudes.
Therefore, since sideband resonances disperse differently with angles,
we can observe multiple resonant features in the noise corresponding to a single dip in conductance, see Figs.~\ref{fig:fano}(a)-(b).

Looking at angle-resolved noise maps we can confirm that resonance peaks in noise are caused by the same inelastic scattering mechanism as for conductance both for double-barrier tunnelling (Fig.~\ref{fig:double}) and Fano (Fig.~\ref{fig:fano}) resonances. One particular feature common to all angle-resolved noise maps is complete suppression of noise for the $\varphi=0$ channel. An analytic derivation, see Appendix~\ref{App_B}, proves that Klein tunnelling is responsible for this effect. There is no reflection at any of sideband energies and all transmitted waves acquire a trivial phase while their amplitudes are modulated as cylindrical harmonics (Bessel functions).
This leads to a sum rule which ensures vanishing shot noise for perpendicular incidence.

\subsection{THz radiation detection}

\begin{figure}[tb!]
\includegraphics[width=\columnwidth]{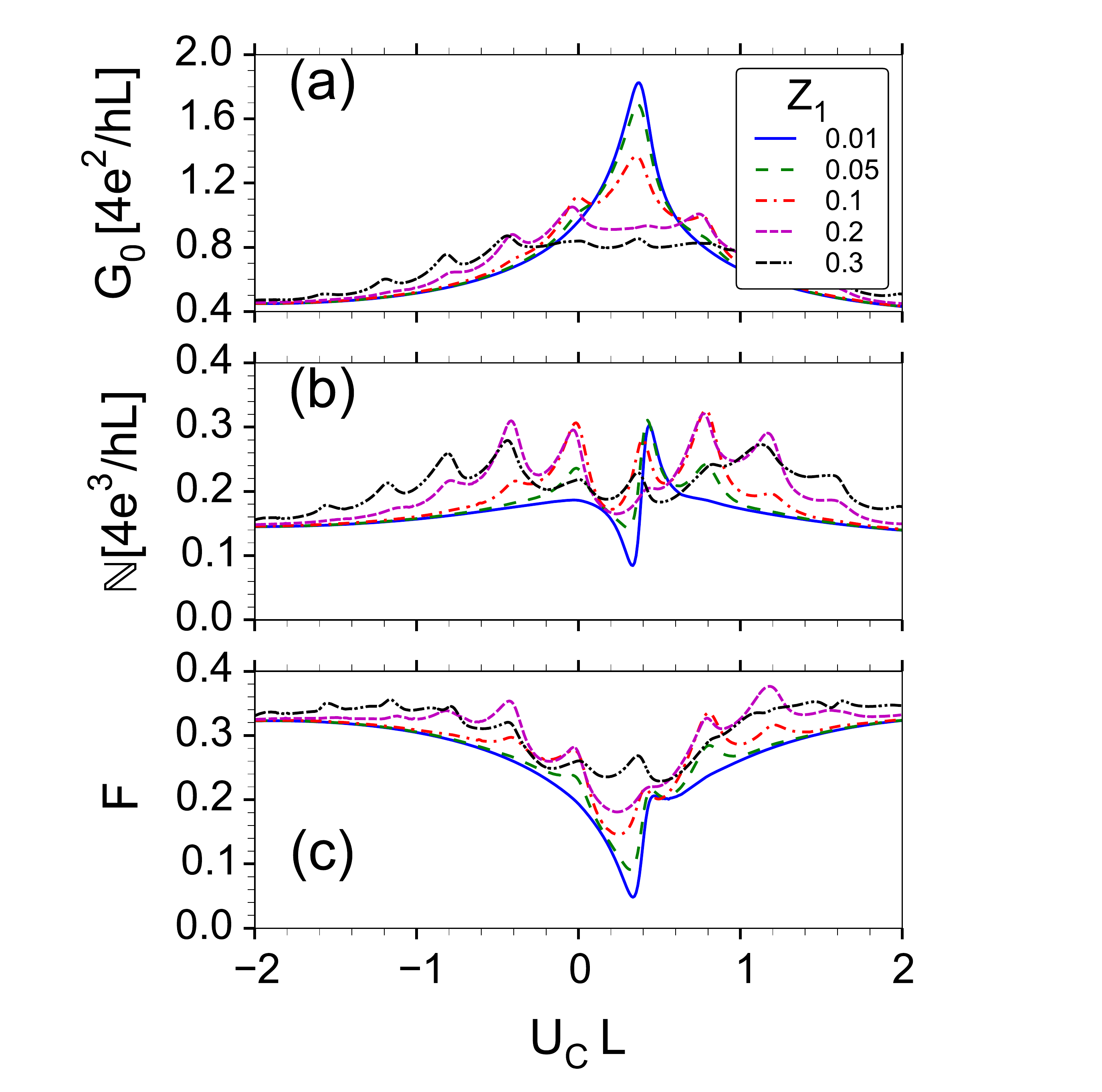}
\caption{Channel doping dependence of detector response. The model parameters are $Z_0=0.48\pi$, $U=-10/L$, and $\Omega=0.4/L$}
\label{fig:detector1}
\end{figure}

Since the static conductance component experiences resonances in the strong contact doping regime even for a relatively weak ac drive strength,
we proposed\cite{Korniyenko:2016hg} that the device in this regime can be used as a THz frequency detector.
In this section we analyze the device's noise characteristics under similar parameters.
As was established in the previous section, both conductance and noise experience quasi-periodic resonance combs, see Fig.~\ref{fig:double}(a)-(b).
As the driving strength is increased, the main resonance peak gets reduced as the sideband peaks are enhanced, see Fig.~\ref{fig:detector1}(a)-(b).
It is therefore quite natural that the corresponding Fano factor is increasing
and approaches the value of $1/3$ observed in the static case in absence of resonance, see Fig.~\ref{fig:detector1}(c)

\begin{figure}[tb!]
\includegraphics[width=\columnwidth]{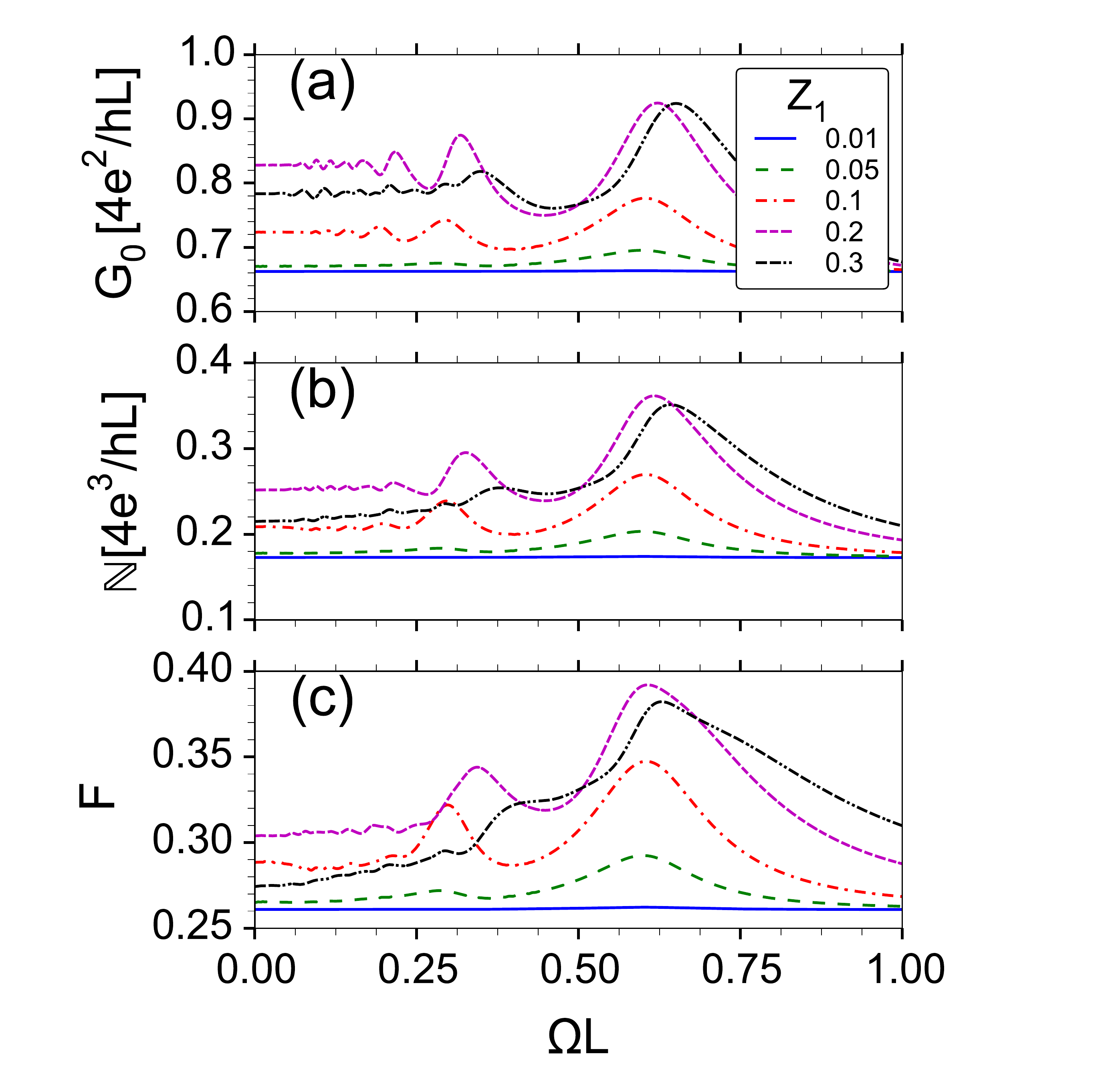}
\caption{Frequency response of the detector. The model parameters are $Z_0=0.48\pi$, $U=-10/L$, and $U_C=1/L$.}
\label{fig:detector2}
\end{figure}

It is also instructive to look at the frequency response of the detector for a fixed value of channel doping, see Fig.~\ref{fig:detector2}.
The conductance displays peaks whenever sideband scattering is done via a quasibound state $E_b=n\Omega$.
Since the resonances manifest themselves in increased noise on resonance, its shape is very similar to that of conductance with a series of resonance peaks.
We note however that the Fano factor is lower for secondary peaks, compared with the main $n=1$ peak, and so is their width.
Thus in an experimental setup a more sensitive narrow bandwidth detector might rely on secondary resonances.
For weak signals, the noise is reduced well below the subpoissonian value, $F<1/3$, which is favorable for a detector's signal-to-noise ratio.

\subsection{Strong driving}

\begin{figure}[tb!]
\includegraphics[width=\columnwidth]{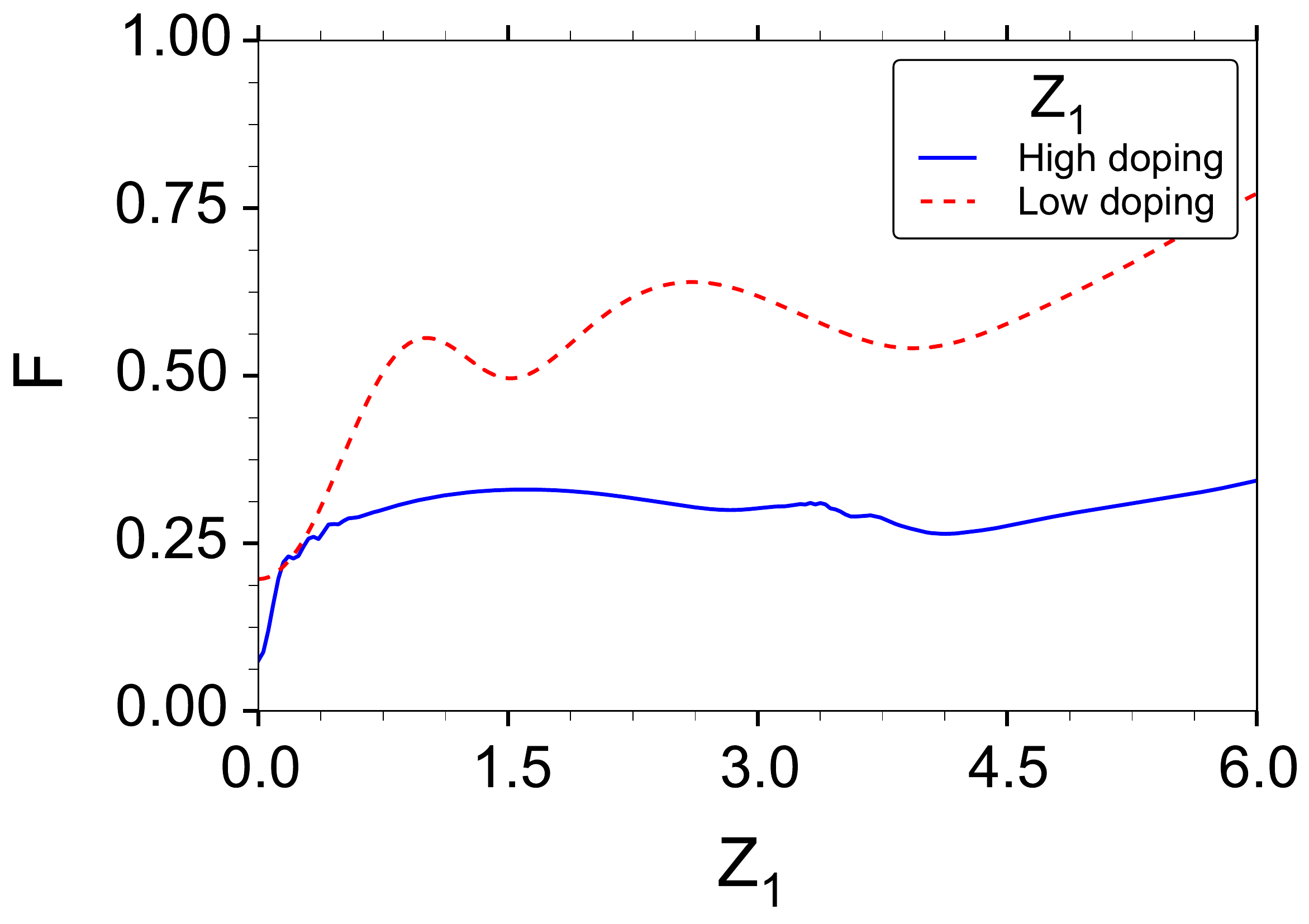}
\caption{Noise performance for high doping (solid) and low doping (dashed) of contacts as function of ac drive strength $Z_1$.
For high doping the parameters are $Z_0=0.4\pi$, $U=-8/L$, $\Omega=0.2/L$, $U_C=1/L$;
for low doping they are $Z_0=0.4\pi$, $U=1.2/L$, $\Omega=1/L$, $U_C=0.75/L$.}
\label{fig:strong}
\end{figure}

In our previous paper we showed that ac conductance harmonics can be selectively enhanced for both double barrier tunnelling and Fano resonance regimes,
thereby allowing the device to be operated as a frequency multiplier.
To generate higher harmonics we need to go to strong driving regime ($Z_1>1$).
Since in-depth physics analysis has been presented in the previous sections,
here we show only a comparative study of the device performance for parameters corresponding to the
resonances in the two regimes (high or low contact doping), see Fig.~\ref{fig:strong}.
Although the zero-frequency noise is not necessarily a figure of merit for harmonic generation, it serves as an indicator of the general device performance.
It is striking that the device at high contact doping, or double tunnelling regime, performs consistently better in terms of shot noise.
Since almost all channels but the resonant ones are closed, the Fano factor remains consistently close to $1/3$ value.
In contrast, at low doping, or Fano/Breit-Wigner resonance regime, most of the channels are highly transparent
and all scattering trajectories contribute to the final result, ramping up the noise as the number of sidebands is increased for stronger drive.

\section{Summary}\label{sec:summary}

We have presented results for the zero frequency shot noise in a ballistic graphene transistor driven by an ac gate potential.
We have analysed two different setups of the transistor potential landscape: high and low doping of electrodes (the parameter $U$).
For dc operation, in the high electrode doping regime, the Fano factor maximum is at $1/3$ for zero top-gate potential ($Z_0=0$)
and is largely characterized by Fabry-P\'erot interferences, in analogy to the conductance and in agreement with the literature.
A non-zero static top-gate potential ($Z_0\ne 0$) allows excitation of a quasibound state in the top gate,
which leads to resonant enhancement of conductance and suppression of noise at that energy.
Including harmonic ac drive we have observed formation of multiple sideband resonances in the noise related to the excitation of the quasibound state.
For high contact doping, the additional resonant transport channel leads to an enhanced Fano factor compared with dc,
but $F<1/3$ also for strong ac drive ($Z_1>1$).
For low contact doping, the resonances in noise follow closely the Fano- and Breit-Wigner resonances also present in the conductance.
We have discussed possible utilization of the device in the high doping regime as a high-frequency radiation detector.
Our results indicate that secondary peak detection could be more experimentally desirable due to higher signal-to-noise ratio.
We have compared the shot noise behavior of a frequency multiplier in two different contact doping regimes for increasing ac driving strength ($Z_1>1$)
and concluded that the high doping device performs better and operates under the diffusive metal limit of $1/3$.
Finally, we have found analytically that Klein tunneling at perpendicular incidence persists under ac drive, which leads to a completely noiseless channel.

\acknowledgments

We acknowledge financial support from the Swedish foundation for strategic research, SSF, and Knut and Alice Wallenberg foundation, KAW.
The research of O.S. was partly supported by the National Science Foundation (Grant No. DMR-1508730).

\begin{appendix}

\section{Expression for the noise and Fano factor}\label{App_A}

In this Appendix we derive expressions for the zero-frequency noise within a Landauer-B\"uttiker scattering approach to transport
and Floquet theory for scattering states,
properly modified to take into account that electrons in graphene are massless Dirac fermions
obeying Eq.~(\ref{Dirac_eq}) with the Hamiltonian in Eq.~(\ref{Hamiltonian}).
We have presented solutions to these equations in our recent papers\cite{Korniyenko:2016ct,Korniyenko:2016hg}.
Below we utilize these solutions to derive expressions for the noise and the differential Fano factor.
Although we have tried to make this appendix self-contained, we emphasize that
it builds on the results in Refs.~\onlinecite{Korniyenko:2016ct,Korniyenko:2016hg}.

\subsection{Scattering basis}\label{app_basis}

Within scattering theory\cite{2000PhR...336....1B} one derives a scattering matrix for the device region connecting incoming and outgoing waves in the leads.
For the scattering matrix to be unitary, a scattering basis is needed. The elementary waves are labeled by the energy $E$ and
the transverse momentum $k_y$ for assumed invariance in the transverse $y$-direction.
For the coordinate system chosen as in the main text we obtain \cite{Korniyenko:2016ct}
\begin{equation}
\begin{aligned}
&\psi_{\rightarrow}(x,k_y,E) = \frac{1}{\sqrt{2\mathrm{v}(k_y,E)}}\!
\begin{pmatrix}
1\\
\eta(k_y,E)
\end{pmatrix}
\!e^{i\kappa(k_y,E)x},\\
&\psi_{\leftarrow}(x,k_y,E) = \frac{1}{\sqrt{2\mathrm{v}(k_y,E)}}\!
\begin{pmatrix}
1\\
\bar{\eta}(k_y,E)
\end{pmatrix}
\!e^{-i\kappa(k_y,E)x},
\end{aligned}
\label{ScatBasis}
\end{equation}
where arrows indicate the direction of propagation along the $x$-axis, and
\begin{equation}
\begin{aligned}
&\eta(k_y,E) = \frac{\kappa(k_y,E)+ik_y}{E},\\
&\bar{\eta}(k_y,E) = \frac{-\kappa(k_y,E)+ik_y}{E},\\
&\mathrm{v}(k_y,E) = \frac{\kappa(k_y,E)}{E},\\
&\kappa(k_y,E) = \mathrm{sgn}(E)\sqrt{E^2-k_y^2}.
\end{aligned}
\label{velocity}
\end{equation}
The normalization of these plane waves is chosen such that they carry unit probability flux along the $x$-axis.
The probability flux is defined as
\begin{equation}
j_x(x,k_y,E) = \psi^{\dagger}(x,k_y,E)\sigma_x\psi(x,k_y,E),
\label{FluxDef}
\end{equation}
and we have $j_x^{\rightarrow}=1$ and $j_x^{\leftarrow}=-1$.

\subsection{Current operator}

The scattering basis introduced in the previous section, see Eq.~(\ref{ScatBasis}),
allows us to define the field operator for quasiparticles in contact $\alpha\in\left\{\mathrm{S,D}\right\}$.
For the two-terminal setup considered in the main text, where the coordinate system is uniquely fixed everywhere,
one can deduce the following expression for the field operator in the drain contact
\begin{align}
&\hat{\Psi}_{\mathrm{D}}(x,y,t) = \sum_{k_y}\frac{e^{ik_yy}}{\sqrt{W_y}}\int\limits_{|E|>|k_y|}\frac{dE}{\sqrt{2\pi}}e^{-iEt}\label{FieldOp}\\
&\!\times\!\left[\hat{\gamma}_{\mathrm{D,in}}(k_y,E)\psi_{\leftarrow}(x,k_y,E)+\hat{\gamma}_{\mathrm{D,out}}(k_y,E)\psi_{\rightarrow}(x,k_y,E)\right]\!,\notag
\end{align}
where $W_y$ is the width in the transverse $y-$direction. Note that we below will utilize periodic boundary conditions in the transverse direction
and the dependence on $W_y$ drops out.
Integration over energy is restricted to quasiparticle states which describe propagating waves.
The operators $\hat{\gamma}_{\mathrm{\alpha,in/out}}(k_y,E)$ annihilate the corresponding incoming/outgoing quasiparticle
with energy $E$ and transverse momentum $k_y$ in the contact $\alpha$, and satisfy the usual fermionic anti-commutation relations
\begin{equation}
\begin{aligned}
\left\{\hat{\gamma}_{\mathrm{\alpha,in}}(k_y,E),\hat{\gamma}^{\dagger}_{\mathrm{\beta,in}}(k^{\prime}_y,E^{\prime})\right\}
&= \delta_{\alpha,\beta}\delta_{k_y,k^{\prime}_y}\delta(E-E^{\prime}),\\
\left\{\hat{\gamma}_{\mathrm{\alpha,in}}(k_y,E),\hat{\gamma}_{\mathrm{\beta,in}}(k^{\prime}_y,E^{\prime})\right\} &= 0,\\
\left\{\hat{\gamma}^{\dagger}_{\mathrm{\alpha,in}}(k_y,E),\hat{\gamma}^{\dagger}_{\mathrm{\beta,in}}(k^{\prime}_y,E^{\prime})\right\} &= 0,
\end{aligned}
\label{CommRels}
\end{equation}
where the contact indices $\alpha,\beta = \{\mathrm{S,D}\}$.
According to scattering theory, the outgoing operator $\hat{\gamma}_{\mathrm{D,out}}(k_y,E)$ is related to the incoming one via a scattering matrix.
For the case of an oscillating barrier and static bias between source and drain contacts this relation reads
\begin{align}
\hat{\gamma}_{\mathrm{D,out}}(k_y,E)
\!= \!\sum\limits_{\beta=\mathrm{S,D}}\sum\limits_{n,\mathrm{prop.}}\!\!S_{\mathrm{D\beta}}(k_y;E,E_n)\hat{\gamma}_{\mathrm{\beta,in}}(k_y,E_n),
\label{GammaOut}
\end{align}
where $E_n = E + n\Omega$. We restrict the sum over side-bands to propagating waves only,
which is equivalent to setting the scattering matrix elements to zero if an incoming/outgoing wave is evanescent.
Now we use the field operator, Eq.~(\ref{FieldOp}), and its Hermitian conjugate to construct the current operator in the drain contact [compare to Eq.~(\ref{FluxDef})],
\begin{align}
\hat{I}_{\mathrm{D}}(x,t) = e\int\limits_{0}^{W_y}\! dy\, \hat{\Psi}^{\dagger}_{\mathrm{D}}(x,y,t)\sigma_x\hat{\Psi}_{\mathrm{D}}(x,y,t),
\label{CurrOp}
\end{align}
where $e$ is the electron charge. The final expression for the current operator in terms of the creation/annihilation operators of the incoming and outgoing quasiparticles has the form
\begin{widetext}
\begin{align}
&\hat{I}_{\mathrm{D}}(x,t) = \frac{e}{2\pi}\sum_{k_y}\int\limits_{|E|>|k_y|}dE\int\limits_{|E'|>|k_y|}dE' e^{i(E-E')t}
\Bigl[
\psi_{\leftarrow}^{\dagger}(x,k_y,E)\sigma_x\psi_{\leftarrow}(x,k_y,E')\hat{\gamma}^{\dagger}_{\mathrm{D,in}}(k_y,E)\hat{\gamma}_{\mathrm{D,in}}(k_y,E')\notag\\
&+\psi_{\leftarrow}^{\dagger}(x,k_y,E)\sigma_x\psi_{\rightarrow}(x,k_y,E')\hat{\gamma}^{\dagger}_{\mathrm{D,in}}(k_y,E)\hat{\gamma}_{\mathrm{D,out}}(k_y,E')
+\psi_{\rightarrow}^{\dagger}(x,k_y,E)\sigma_x\psi_{\leftarrow}(x,k_y,E')\hat{\gamma}^{\dagger}_{\mathrm{D,out}}(k_y,E)\hat{\gamma}_{\mathrm{D,in}}(k_y,E')\notag\\
&+\psi_{\rightarrow}^{\dagger}(x,k_y,E)\sigma_x\psi_{\rightarrow}(x,k_y,E')\hat{\gamma}^{\dagger}_{\mathrm{D,out}}(k_y,E)\hat{\gamma}_{\mathrm{D,out}}(k_y,E')
\Bigr].
\label{CurrOp2}
\end{align}
\end{widetext}

\subsection{Noise formulas}

By definition, current noise is a matrix of correlation functions between currents in the contacts of the system,
with matrix elements given by \cite{1998PhRvB..5812993P,2000PhR...336....1B}
\begin{align}
\mathcal{N}_{\alpha\beta}(\tau) = \frac{1}{2\mathcal{T}}\int\limits_{0}^{\mathcal{T}}dt\left\langle\left\{
\Delta\hat{I}_{\alpha}(x,t+\tau),\Delta\hat{I}_{\beta}(x,t)\right\}\right\rangle.
\label{NoiseDef}
\end{align}
Here, $\Delta\hat{I}_{\alpha}(x,t) = \hat{I}_{\alpha}(x,t) - \langle\hat{I}_{\alpha}(x,t)\rangle$ is the deviation of
the current operator in contact $\alpha$ from its mean value, where $\langle\cdot\rangle$ means statistical average.
Taking into account that there is an oscillating perturbation in our setup (ac-driven gate),
we also average the noise over one oscillation period $\mathcal{T} = 2\pi/\Omega$.
In many experiments, the quantity of primary interest is the zero-frequency noise, which is obtained from Eq.~(\ref{NoiseDef}) via
\begin{equation}
N_{\alpha\beta}(0) = \int\limits_{-\infty}^{\infty}d\tau\mathcal{N}_{\alpha\beta}(\tau).
\end{equation}
Below we will only consider the zero-frequency noise and therefore we will omit its argument, $N_{\alpha\beta}\equiv N_{\alpha\beta}(0)$.
It can be shown \cite{1998PhRvB..5812993P} that zero-frequency noise satisfies the conservation law
\begin{equation}
\sum\limits_{\alpha}N_{\alpha\beta} = \sum\limits_{\beta}N_{\alpha\beta} = 0,
\end{equation}
which allows us to consider only $N = N_{\mathrm{DD}} = N_{\mathrm{SS}}$.
The cross correlations $N_{\mathrm{SD}} = N_{\mathrm{DS}} = -N_{\mathrm{DD}}$.
By substituting Eq.~(\ref{CurrOp2}) into Eq.~(\ref{NoiseDef}), we can derive the noise formula.
Every operator for an outgoing quasiparticle must be expressed in terms of the corresponding operators for the incoming ones, using Eq.~(\ref{GammaOut}).
Whenever one encounters a statistical average of a product of four creation/annihilation operators, one can simplify it using Wick's theorem:
\begin{align}
&\langle\hat{\gamma}^{\dagger}_{\mathrm{\alpha,in}}(k_{y},E_1)\hat{\gamma}_{\mathrm{\beta,in}}(k_{y},E_1^{\prime})\hat{\gamma}^{\dagger}_{\mathrm{\delta,in}}(k_{y}^{\prime},E_2)\hat{\gamma}_{\mathrm{\lambda,in}}(k_{y}^{\prime},E_2^{\prime})\rangle\notag\\
&\!=\!\langle\hat{\gamma}^{\dagger}_{\mathrm{\alpha,in}}(k_{y},E_1)\hat{\gamma}_{\mathrm{\beta,in}}(k_{y},E_1^{\prime})\rangle
\langle\hat{\gamma}^{\dagger}_{\mathrm{\delta,in}}(k_{y}^{\prime},E_2)\hat{\gamma}_{\mathrm{\lambda,in}}(k_{y}^{\prime},E_2^{\prime})\rangle\notag\\
&\!+\!\langle\hat{\gamma}^{\dagger}_{\mathrm{\alpha,in}}(k_{y},E_1)\hat{\gamma}_{\mathrm{\lambda,in}}(k_{y}^{\prime},E_2^{\prime})\rangle
\langle\hat{\gamma}_{\mathrm{\beta,in}}(k_{y},E_1^{\prime})\hat{\gamma}^{\dagger}_{\mathrm{\delta,in}}(k_{y}^{\prime},E_2)\rangle.
\end{align}
Finally, statistical averaging is performed assuming that the contacts of the system are kept at local equilibrium,
\begin{align}
&\langle\hat{\gamma}^{\dagger}_{\mathrm{\alpha,in}}(k_{y},E)\hat{\gamma}_{\mathrm{\beta,in}}(k_{y}^{\prime},E^{\prime})\rangle \notag\\
&\phantom{\langle\hat{\gamma}^{\dagger}_{\mathrm{\alpha,in}}(k_{y},E)\hat{\gamma}}= \delta_{\alpha,\beta}\delta_{k_{y},k_{y}^{\prime}}\delta(E-E^{\prime})f_{\alpha}(E),\\
&\langle\hat{\gamma}_{\mathrm{\alpha,in}}(k_{y},E)\hat{\gamma}^{\dagger}_{\mathrm{\beta,in}}(k_{y}^{\prime},E^{\prime})\rangle \notag\\
&\phantom{\langle\hat{\gamma}^{\dagger}_{\mathrm{\alpha,in}}(k_{y},E)\hat{\gamma}}= \delta_{\alpha,\beta}\delta_{k_{y},k_{y}^{\prime}}\delta(E-E^{\prime})[1 - f_{\alpha}(E)],
\end{align}
where $f_{\alpha}(E)$ is the Fermi-Dirac distribution in contact $\alpha$.
Performing this lengthy but straightforward calculation, one can obtain a rather compact formula for the noise $N$, which can be written as \cite{Moskalets:2004ct}
\begin{align}
N = N_{\mathrm{th}} + N_{\mathrm{sh}},
\end{align}
where $N_{\mathrm{th}}$ is the thermal or Johnson-Nyquist noise, and $N_{\mathrm{sh}}$ is the shot noise.
Thermal noise is expressed by the formula
\begin{widetext}
\begin{align}
N_{\mathrm{th}}= \frac{e^2}{2\pi}\sum\limits_{\beta=\mathrm{S,D}}\sum\limits_{l=-\infty}^{\infty}
\sum_{k_y}\int\limits_{\mathrm{prop.}}dE
\left\{\delta_{\alpha,D}\left[\delta_{l,0} - 2\left|S_{\mathrm{DD}}(k_y;E_l,E)\right|^2\right] + \left|S_{\mathrm{D\alpha}}(k_y;E_l,E)\right|^2\right\}\!f_{\alpha}(E)[1-f_{\alpha}(E)],
\label{NoiseTherm}
\end{align}
%
where integration over energy runs only over propagating states, i.e. $|E|>|k_y|$ and $|E_l|>|k_y|$.
We note that for temperatures $T\rightarrow 0$, the combination of Fermi functions appearing in Eq.~(\ref{NoiseTherm}) leads to $N_{\mathrm{th}}\rightarrow 0$. Therefore, at low temperatures one can neglect thermal noise and focus on the shot noise. The latter can be written in the form
\begin{align}
N_{\mathrm{sh}}= \frac{e^2}{2\pi}\sum_{\alpha,\beta=\mathrm{S,D}}\sum\limits_{l,n,m=-\infty}^{\infty}
\sum_{k_y}\int\limits_{\mathrm{prop.}}dE\,
\frac{[f_{\alpha}(E_l) - f_{\beta}(E_m)]^2}{2}
\left[S_{\mathrm{D\alpha}}(k_y;E,E_l)\right]^{\dagger}S_{\mathrm{D\alpha}}(k_y;E_n,E_l)\notag\\
\times S_{\mathrm{D\beta}}(k_y;E,E_m)\left[S_{\mathrm{D\beta}}(k_y;E_n,E_m)\right]^{\dagger}.
\label{NoiseShot}
\end{align}
We note that there is a complex conjugation symmetry in the kernel under index interchange $l\leftrightarrow m$.
This means that shot noise $N_{\mathrm{sh}}$ is a purely real quantity.
\end{widetext}

\subsection{Shot noise formula at zero temperature}

At zero temperature the Fermi function factor in Eq.~(\ref{NoiseShot}) simplifies to step functions.
We are interested in the linear response to the applied source-drain bias voltage $V_{\mathrm{S}}$ (we set $V_{\mathrm{D}}=0$) and it is instructive to calculate the differential noise
\be
\mathbb{N} = \left.\frac{\partial N_{\mathrm{sh}}}{\partial V_{\mathrm{S}}}\right|_{V_{\mathrm{S}}\rightarrow 0}.
\ee
The terms with the Fermi function of the drain give zero contribution while those of the source are reduced to a delta function
\be
\left.\frac{\partial \theta(E_l-E_F+eV_{\mathrm{S}})}{\partial V_{\mathrm{S}}}\right|_{V_{\mathrm{S}}\rightarrow 0}=e\delta(E_l-E_F).
\ee
The integral over energy can be shifted so that the differential noise kernel is written only in terms of two independent scattering states (sideband "ladders").
The two scattering states have two different quasienergies (energies of the incoming/outgoing waves in the leads), one at
the Fermi energy $E_F$ and one shifted away from it by $(m-l)\Omega$. The zero-temperature shot noise is then reduced to
\begin{widetext}
\be
\mathbb{N}=\frac{e^3}{2\pi h}\sum_{k_y}\sum_{lqn}\mathrm{Re}\Big\{t^*_{l}(E_F)t_{n+l}(E_F)\Big(r'_{l-q}(E_F+q\Omega)r'^*_{n+l-q}(E_F+q\Omega)+(1-\delta_{q0})t_{l-q}(E_F+q\Omega)t^*_{n+l-q}(E_F+q\Omega)\Big)\Big\},
\label{eq:Nfinal}
\ee
\end{widetext}
where $q=m-l$. Here we explicitly write the scattering matrix elements as reflection and transmission coefficients.
Note that unprimed quantities $r_n(E)$ and $t_n(E)$ [primed quantities $r'_n(E)$ and $t'_n(E)$] are obtained for an incident wave
from the source (drain) at energy $E$, scattered to an energy $E_n$.
%

\subsection{Differential Fano factor}

We define the differential Fano factor as the ratio between the differential noise and the dc conductance, i.e.
\begin{equation}
F = \left.\frac{\partial N_{\mathrm{sh}}/\partial V_{\mathrm{S}}}
{e\partial I_{\mathrm{D}}/\partial V_{\mathrm{S}}}\right|_{V_{\mathrm{S}}\rightarrow 0},
\end{equation}
where $N_{\mathrm{sh}}$ is given by Eq.~(\ref{NoiseShot})
and the dc component of the drain current $I_{\mathrm{D}}$ is obtained by averaging the current operator in Eq.~(\ref{CurrOp2}) as
\begin{align}
I_{\mathrm{D}} = \frac{1}{\mathcal{T}}\int\limits_{0}^{\mathcal{T}}dt
\langle\hat{I}_{\mathrm{D}}(x,t)\rangle.
\end{align}
The result has the form
\begin{align}
&\!\!\!I_{\mathrm{D}} = \frac{e}{2\pi}\sum_{k_y}\int\limits_{\mathrm{prop.}}dE
\sum\limits_{\alpha=\mathrm{S,D}}\sum_{n=-\infty}^{\infty}\notag\\
&\!\!\!\left[S_{\mathrm{D\alpha}}(k_y;E_n,E)\right]^{\dagger}S_{\mathrm{D\alpha}}(k_y;E_n,E)
[f_{\alpha}(E) - f_{\mathrm{D}}(E_n)].
\end{align}
At zero temperature, we get a Fano factor
\begin{equation}
F = \frac{\mathbb{N}}{eG_0}
\label{eq:F}
\end{equation}
where $\mathbb{N}$ is given by Eq.~(\ref{eq:Nfinal}) and the zero temperature dc conductance $G_0$ is given as Eq.~(7) in Ref.~\onlinecite{Korniyenko:2016hg}.

\section{Noiseless inelastic Klein tunnelling for $k_y=0$}\label{App_B}

The reflection coefficient in the static case vanishes during tunnelling at perpendicular incidence to the barrier $k_y=0$ (Klein tunneling).
Here we take it one step further and show that the same holds for all sidebands for photon-assisted tunnelling, which leads to a noiseless quantum channel.
The reflection and transmission coefficients are determined by the following equations:
\bea
\label{eq:bc}
r_n=\sum_m\vec C^{\,T}_n\check M_{nm}\vec B_m t_m,\label{r_n} \nonumber\\
\sum_m \vec A^{\,T}_n \check M_{nm}\vec B_m t_m=\delta_{n0}.\label{t_n}
\eea
The matrix $\check{M}_{nm} = \exp[iZ_0\sigma_x](i\sigma_x)^{|n-m|}J_{|n-m|}(Z_1)$, where $J_n$ is the $n-$th Bessel function of the first kind.
We studied this matrix in detail in Ref.~\onlinecite{Korniyenko:2016hg}.
There we also gave the expressions for the pseudospin vectors $\vec A_n$, $\vec B_n$, and $\vec C_n$ as Eqs.~(B12)-(B14), and we do not repeat them here.
The important point here is that for perpendicular incidence these vectors reduce to a very simple form:
\bea
\vec A_n &=& \left[ \spin(1,1) e^{-i\kappa_n L_1}
                           \right]
{e^{i\kappa_n^LL_1}} ,\\
\vec B_n &=& \left[  \spin(1,1)       e^{-i\kappa_n L_2}
                            \right]
e^{i\kappa_n^R L_2},\\
\vec C_n &=& \left[ \spin(1,-1)        e^{i\kappa_n L_1} \right]
e^{-i\kappa_n^L L_1} ,
\eea
where $\kappa_n^{L/R}=\kappa(k_y,E_n-U_{L/R})$, c.f. Eq.~(\ref{velocity}).
Due to the peculiar pseudospin structure, the product $\vec{C}_n^T\check{M}_{nm}\vec{B}_m$  in Eq.~(\ref{eq:bc}) vanishes,
thus leading to $r_n\equiv0$, $\forall n$.
In other words, Klein tunnelling implies no backscattering even from an ac-driven delta barrier.
The particles are allowed to scatter between energy sidebands but the barrier remains effectively transparent.
The barrier now contributes a trivial phase to all sideband channels and the matrix in Eq.~(\ref{eq:bc}) for transmission coefficients is easily inverted giving
\be
t_m=\frac{1}{2}\exp(-iZ_0)(-i)^{|m|}J_{|m|}(Z_1)e^{i(U_L L_1+U_R L_2)}
\ee
Note that the transmission amplitudes for this delta barrier are energy-independent. 
The differential noise kernel in Eq.~(\ref{eq:Nfinal}) at $k_y=0$ is now expressed only in terms of such transmission functions,
\begin{align}
&\sum_{ln}\sum_{q\neq0}\mathrm{Re}\Big\{t^*_{l}t_{n+l}t_{l-q}t_{n+l-q}^*\Big\} \nonumber\\
&\propto \sum_{lp}\sum_{q\neq0}\mathrm{Re}\Big\{\frac{i^{|l|+|p+q|}}{i^{|p|+|l+q|}}J_{|l|}J_{|p|}J_{|l+q|}J_{|p+q|}\Big\} \label{cancelation}\\
&=0,\nonumber
\end{align}
where we used a short hand notation such that all Bessel functions should be evaluated as $J_n=J_n(Z_1)$.
In the second line in Eq.~(\ref{cancelation}) we set $p=n+l$ and in the last step we used the following orthonormal property of Bessel functions
\be
\sum_p (i^{|p+q|-|p|})J_{|p|} J_{|p+q|}=\delta_{q0}.
\ee
With that we have proven that the differential noise kernel at $k_y=0$ vanishes. 

\end{appendix}


\begin{thebibliography}{61}%
\makeatletter
\providecommand \@ifxundefined [1]{%
 \@ifx{#1\undefined}
}%
\providecommand \@ifnum [1]{%
 \ifnum #1\expandafter \@firstoftwo
 \else \expandafter \@secondoftwo
 \fi
}%
\providecommand \@ifx [1]{%
 \ifx #1\expandafter \@firstoftwo
 \else \expandafter \@secondoftwo
 \fi
}%
\providecommand \natexlab [1]{#1}%
\providecommand \enquote  [1]{``#1''}%
\providecommand \bibnamefont  [1]{#1}%
\providecommand \bibfnamefont [1]{#1}%
\providecommand \citenamefont [1]{#1}%
\providecommand \href@noop [0]{\@secondoftwo}%
\providecommand \href [0]{\begingroup \@sanitize@url \@href}%
\providecommand \@href[1]{\@@startlink{#1}\@@href}%
\providecommand \@@href[1]{\endgroup#1\@@endlink}%
\providecommand \@sanitize@url [0]{\catcode `\\12\catcode `\$12\catcode
  `\&12\catcode `\#12\catcode `\^12\catcode `\_12\catcode `\%12\relax}%
\providecommand \@@startlink[1]{}%
\providecommand \@@endlink[0]{}%
\providecommand \url  [0]{\begingroup\@sanitize@url \@url }%
\providecommand \@url [1]{\endgroup\@href {#1}{\urlprefix }}%
\providecommand \urlprefix  [0]{URL }%
\providecommand \Eprint [0]{\href }%
\providecommand \doibase [0]{http://dx.doi.org/}%
\providecommand \selectlanguage [0]{\@gobble}%
\providecommand \bibinfo  [0]{\@secondoftwo}%
\providecommand \bibfield  [0]{\@secondoftwo}%
\providecommand \translation [1]{[#1]}%
\providecommand \BibitemOpen [0]{}%
\providecommand \bibitemStop [0]{}%
\providecommand \bibitemNoStop [0]{.\EOS\space}%
\providecommand \EOS [0]{\spacefactor3000\relax}%
\providecommand \BibitemShut  [1]{\csname bibitem#1\endcsname}%
\let\auto@bib@innerbib\@empty
\bibitem [{\citenamefont {Castro~Neto}\ \emph {et~al.}(2009)\citenamefont
  {Castro~Neto}, \citenamefont {Guinea}, \citenamefont {Peres}, \citenamefont
  {Novoselov},\ and\ \citenamefont {Geim}}]{CastroNeto:2009cl}%
  \BibitemOpen
  \bibfield  {author} {\bibinfo {author} {\bibfnamefont {A.~H.}\ \bibnamefont
  {Castro~Neto}}, \bibinfo {author} {\bibfnamefont {F.}~\bibnamefont {Guinea}},
  \bibinfo {author} {\bibfnamefont {N.~M.~R.}\ \bibnamefont {Peres}}, \bibinfo
  {author} {\bibfnamefont {K.~S.}\ \bibnamefont {Novoselov}}, \ and\ \bibinfo
  {author} {\bibfnamefont {A.~K.}\ \bibnamefont {Geim}},\ }\bibfield  {title}
  {\enquote {\bibinfo {title} {{The electronic properties of graphene}},}\
  }\href@noop {} {\bibfield  {journal} {\bibinfo  {journal} {Reviews Of Modern
  Physics}\ }\textbf {\bibinfo {volume} {81}},\ \bibinfo {pages} {109--162}
  (\bibinfo {year} {2009})}\BibitemShut {NoStop}%
\bibitem [{\citenamefont {Peres}(2010)}]{Peres:2010tn}%
  \BibitemOpen
  \bibfield  {author} {\bibinfo {author} {\bibfnamefont {N.}~\bibnamefont
  {Peres}},\ }\bibfield  {title} {\enquote {\bibinfo {title} {{Colloquium: the
  transport properties of graphene: an introduction}},}\ }\href@noop {}
  {\bibfield  {journal} {\bibinfo  {journal} {Reviews Of Modern Physics}\
  }\textbf {\bibinfo {volume} {82}},\ \bibinfo {pages} {2673} (\bibinfo {year}
  {2010})}\BibitemShut {NoStop}%
\bibitem [{\citenamefont {Novoselov}\ \emph {et~al.}(2012)\citenamefont
  {Novoselov}, \citenamefont {ko}, \citenamefont {Colombo}, \citenamefont
  {Gellert}, \citenamefont {Schwab},\ and\ \citenamefont
  {Kim}}]{Novoselov:2012hw}%
  \BibitemOpen
  \bibfield  {author} {\bibinfo {author} {\bibfnamefont {K.~S.}\ \bibnamefont
  {Novoselov}}, \bibinfo {author} {\bibfnamefont {V.~I.~F.}\ \bibnamefont
  {ko}}, \bibinfo {author} {\bibfnamefont {L.}~\bibnamefont {Colombo}},
  \bibinfo {author} {\bibfnamefont {P.~R.}\ \bibnamefont {Gellert}}, \bibinfo
  {author} {\bibfnamefont {M.~G.}\ \bibnamefont {Schwab}}, \ and\ \bibinfo
  {author} {\bibfnamefont {K.}~\bibnamefont {Kim}},\ }\bibfield  {title}
  {\enquote {\bibinfo {title} {{A roadmap for graphene}},}\ }\href@noop {}
  {\bibfield  {journal} {\bibinfo  {journal} {Nature}\ }\textbf {\bibinfo
  {volume} {490}},\ \bibinfo {pages} {192--200} (\bibinfo {year}
  {2012})}\BibitemShut {NoStop}%
\bibitem [{\citenamefont {Ferrari}\ \emph {et~al.}(2015)\citenamefont
  {Ferrari}, \citenamefont {Bonaccorso}, \citenamefont {Fal'ko}, \citenamefont
  {Novoselov}, \citenamefont {Roche}, \citenamefont {B{\o}ggild}, \citenamefont
  {Borini}, \citenamefont {Koppens}, \citenamefont {Palermo}, \citenamefont
  {Pugno}, \citenamefont {Garrido}, \citenamefont {Sordan}, \citenamefont
  {Bianco}, \citenamefont {Ballerini}, \citenamefont {Prato}, \citenamefont
  {Lidorikis}, \citenamefont {Kivioja}, \citenamefont {Marinelli},
  \citenamefont {Ryh{\"a}nen}, \citenamefont {Morpurgo}, \citenamefont
  {Coleman}, \citenamefont {Nicolosi}, \citenamefont {Colombo}, \citenamefont
  {Fert}, \citenamefont {Garcia-Hernandez}, \citenamefont {Bachtold},
  \citenamefont {Schneider}, \citenamefont {Guinea}, \citenamefont {Dekker},
  \citenamefont {Barbone}, \citenamefont {Sun}, \citenamefont {Galiotis},
  \citenamefont {Grigorenko}, \citenamefont {Konstantatos}, \citenamefont
  {Kis}, \citenamefont {Katsnelson}, \citenamefont {Vandersypen}, \citenamefont
  {Loiseau}, \citenamefont {Morandi}, \citenamefont {Neumaier}, \citenamefont
  {Treossi}, \citenamefont {Pellegrini}, \citenamefont {Polini}, \citenamefont
  {Tredicucci}, \citenamefont {Williams}, \citenamefont {Hong}, \citenamefont
  {Ahn}, \citenamefont {Kim}, \citenamefont {Zirath}, \citenamefont {van Wees},
  \citenamefont {van~der Zant}, \citenamefont {Occhipinti}, \citenamefont
  {Di~Matteo}, \citenamefont {Kinloch}, \citenamefont {Seyller}, \citenamefont
  {Quesnel}, \citenamefont {Feng}, \citenamefont {Teo}, \citenamefont
  {Rupesinghe}, \citenamefont {Hakonen}, \citenamefont {Neil}, \citenamefont
  {Tannock}, \citenamefont {Lofwander},\ and\ \citenamefont
  {Kinaret}}]{2015Nanos...7.4598F}%
  \BibitemOpen
  \bibfield  {author} {\bibinfo {author} {\bibfnamefont {A.~C.}\ \bibnamefont
  {Ferrari}}, \bibinfo {author} {\bibfnamefont {F.}~\bibnamefont {Bonaccorso}},
  \bibinfo {author} {\bibfnamefont {V.}~\bibnamefont {Fal'ko}}, \bibinfo
  {author} {\bibfnamefont {K.~S.}\ \bibnamefont {Novoselov}}, \bibinfo {author}
  {\bibfnamefont {S.}~\bibnamefont {Roche}}, \bibinfo {author} {\bibfnamefont
  {P.}~\bibnamefont {B{\o}ggild}}, \bibinfo {author} {\bibfnamefont
  {S.}~\bibnamefont {Borini}}, \bibinfo {author} {\bibfnamefont {F.~H.~L.}\
  \bibnamefont {Koppens}}, \bibinfo {author} {\bibfnamefont {V.}~\bibnamefont
  {Palermo}}, \bibinfo {author} {\bibfnamefont {N.}~\bibnamefont {Pugno}},
  \bibinfo {author} {\bibfnamefont {J.~A.}\ \bibnamefont {Garrido}}, \bibinfo
  {author} {\bibfnamefont {R.}~\bibnamefont {Sordan}}, \bibinfo {author}
  {\bibfnamefont {A.}~\bibnamefont {Bianco}}, \bibinfo {author} {\bibfnamefont
  {L.}~\bibnamefont {Ballerini}}, \bibinfo {author} {\bibfnamefont
  {M.}~\bibnamefont {Prato}}, \bibinfo {author} {\bibfnamefont
  {E.}~\bibnamefont {Lidorikis}}, \bibinfo {author} {\bibfnamefont
  {J.}~\bibnamefont {Kivioja}}, \bibinfo {author} {\bibfnamefont
  {C.}~\bibnamefont {Marinelli}}, \bibinfo {author} {\bibfnamefont
  {T.}~\bibnamefont {Ryh{\"a}nen}}, \bibinfo {author} {\bibfnamefont
  {A.}~\bibnamefont {Morpurgo}}, \bibinfo {author} {\bibfnamefont {J.~N.}\
  \bibnamefont {Coleman}}, \bibinfo {author} {\bibfnamefont {V.}~\bibnamefont
  {Nicolosi}}, \bibinfo {author} {\bibfnamefont {L.}~\bibnamefont {Colombo}},
  \bibinfo {author} {\bibfnamefont {A.}~\bibnamefont {Fert}}, \bibinfo {author}
  {\bibfnamefont {M.}~\bibnamefont {Garcia-Hernandez}}, \bibinfo {author}
  {\bibfnamefont {A.}~\bibnamefont {Bachtold}}, \bibinfo {author}
  {\bibfnamefont {G.~F.}\ \bibnamefont {Schneider}}, \bibinfo {author}
  {\bibfnamefont {F.}~\bibnamefont {Guinea}}, \bibinfo {author} {\bibfnamefont
  {C.}~\bibnamefont {Dekker}}, \bibinfo {author} {\bibfnamefont
  {M.}~\bibnamefont {Barbone}}, \bibinfo {author} {\bibfnamefont
  {Z.}~\bibnamefont {Sun}}, \bibinfo {author} {\bibfnamefont {C.}~\bibnamefont
  {Galiotis}}, \bibinfo {author} {\bibfnamefont {A.~N.}\ \bibnamefont
  {Grigorenko}}, \bibinfo {author} {\bibfnamefont {G.}~\bibnamefont
  {Konstantatos}}, \bibinfo {author} {\bibfnamefont {A.}~\bibnamefont {Kis}},
  \bibinfo {author} {\bibfnamefont {M.}~\bibnamefont {Katsnelson}}, \bibinfo
  {author} {\bibfnamefont {L.}~\bibnamefont {Vandersypen}}, \bibinfo {author}
  {\bibfnamefont {A.}~\bibnamefont {Loiseau}}, \bibinfo {author} {\bibfnamefont
  {V.}~\bibnamefont {Morandi}}, \bibinfo {author} {\bibfnamefont
  {D.}~\bibnamefont {Neumaier}}, \bibinfo {author} {\bibfnamefont
  {E.}~\bibnamefont {Treossi}}, \bibinfo {author} {\bibfnamefont
  {V.}~\bibnamefont {Pellegrini}}, \bibinfo {author} {\bibfnamefont
  {M.}~\bibnamefont {Polini}}, \bibinfo {author} {\bibfnamefont
  {A.}~\bibnamefont {Tredicucci}}, \bibinfo {author} {\bibfnamefont {G.~M.}\
  \bibnamefont {Williams}}, \bibinfo {author} {\bibfnamefont {B.~H.}\
  \bibnamefont {Hong}}, \bibinfo {author} {\bibfnamefont {J.-H.}\ \bibnamefont
  {Ahn}}, \bibinfo {author} {\bibfnamefont {J.~M.}\ \bibnamefont {Kim}},
  \bibinfo {author} {\bibfnamefont {H.}~\bibnamefont {Zirath}}, \bibinfo
  {author} {\bibfnamefont {B.~J.}\ \bibnamefont {van Wees}}, \bibinfo {author}
  {\bibfnamefont {H.}~\bibnamefont {van~der Zant}}, \bibinfo {author}
  {\bibfnamefont {L.}~\bibnamefont {Occhipinti}}, \bibinfo {author}
  {\bibfnamefont {A.}~\bibnamefont {Di~Matteo}}, \bibinfo {author}
  {\bibfnamefont {I.~A.}\ \bibnamefont {Kinloch}}, \bibinfo {author}
  {\bibfnamefont {T.}~\bibnamefont {Seyller}}, \bibinfo {author} {\bibfnamefont
  {E.}~\bibnamefont {Quesnel}}, \bibinfo {author} {\bibfnamefont
  {X.}~\bibnamefont {Feng}}, \bibinfo {author} {\bibfnamefont {K.}~\bibnamefont
  {Teo}}, \bibinfo {author} {\bibfnamefont {N.}~\bibnamefont {Rupesinghe}},
  \bibinfo {author} {\bibfnamefont {P.}~\bibnamefont {Hakonen}}, \bibinfo
  {author} {\bibfnamefont {S.~R.~T.}\ \bibnamefont {Neil}}, \bibinfo {author}
  {\bibfnamefont {Q.}~\bibnamefont {Tannock}}, \bibinfo {author} {\bibfnamefont
  {T.}~\bibnamefont {Lofwander}}, \ and\ \bibinfo {author} {\bibfnamefont
  {J.}~\bibnamefont {Kinaret}},\ }\bibfield  {title} {\enquote {\bibinfo
  {title} {{Science and technology roadmap for graphene, related
  two-dimensional crystals, and hybrid systems.}}}\ }\href@noop {} {\bibfield
  {journal} {\bibinfo  {journal} {Nanoscale}\ }\textbf {\bibinfo {volume}
  {7}},\ \bibinfo {pages} {4598--4810} (\bibinfo {year} {2015})}\BibitemShut
  {NoStop}%
\bibitem [{\citenamefont {Katsnelson}\ \emph {et~al.}(2006)\citenamefont
  {Katsnelson}, \citenamefont {Novoselov},\ and\ \citenamefont
  {Geim}}]{Katsnelson:2006kd}%
  \BibitemOpen
  \bibfield  {author} {\bibinfo {author} {\bibfnamefont {M.~I.}\ \bibnamefont
  {Katsnelson}}, \bibinfo {author} {\bibfnamefont {K.~S.}\ \bibnamefont
  {Novoselov}}, \ and\ \bibinfo {author} {\bibfnamefont {A.~K.}\ \bibnamefont
  {Geim}},\ }\bibfield  {title} {\enquote {\bibinfo {title} {{Chiral tunnelling
  and the Klein paradox in graphene}},}\ }\href@noop {} {\bibfield  {journal}
  {\bibinfo  {journal} {Nature Physics}\ }\textbf {\bibinfo {volume} {2}},\
  \bibinfo {pages} {620--625} (\bibinfo {year} {2006})}\BibitemShut {NoStop}%
\bibitem [{\citenamefont {Cheianov}\ \emph {et~al.}(2007)\citenamefont
  {Cheianov}, \citenamefont {Fal'ko},\ and\ \citenamefont
  {Altshuler}}]{Cheianov:2007in}%
  \BibitemOpen
  \bibfield  {author} {\bibinfo {author} {\bibfnamefont {V.~V.}\ \bibnamefont
  {Cheianov}}, \bibinfo {author} {\bibfnamefont {V.}~\bibnamefont {Fal'ko}}, \
  and\ \bibinfo {author} {\bibfnamefont {B.~L.}\ \bibnamefont {Altshuler}},\
  }\bibfield  {title} {\enquote {\bibinfo {title} {{The focusing of electron
  flow and a Veselago lens in graphene p-n junctions}},}\ }\href@noop {}
  {\bibfield  {journal} {\bibinfo  {journal} {Science (New York, NY)}\ }\textbf
  {\bibinfo {volume} {315}},\ \bibinfo {pages} {1252--1255} (\bibinfo {year}
  {2007})}\BibitemShut {NoStop}%
\bibitem [{\citenamefont {Novoselov}\ \emph {et~al.}(2005)\citenamefont
  {Novoselov}, \citenamefont {Geim}, \citenamefont {Morozov}, \citenamefont
  {Jiang}, \citenamefont {Katsnelson}, \citenamefont {Grigorieva},
  \citenamefont {Dubonos},\ and\ \citenamefont {Firsov}}]{Novoselov:2005es}%
  \BibitemOpen
  \bibfield  {author} {\bibinfo {author} {\bibfnamefont {K.}~\bibnamefont
  {Novoselov}}, \bibinfo {author} {\bibfnamefont {A.}~\bibnamefont {Geim}},
  \bibinfo {author} {\bibfnamefont {S.}~\bibnamefont {Morozov}}, \bibinfo
  {author} {\bibfnamefont {D.}~\bibnamefont {Jiang}}, \bibinfo {author}
  {\bibfnamefont {M.}~\bibnamefont {Katsnelson}}, \bibinfo {author}
  {\bibfnamefont {I.}~\bibnamefont {Grigorieva}}, \bibinfo {author}
  {\bibfnamefont {S.}~\bibnamefont {Dubonos}}, \ and\ \bibinfo {author}
  {\bibfnamefont {A.}~\bibnamefont {Firsov}},\ }\bibfield  {title} {\enquote
  {\bibinfo {title} {{Two-dimensional gas of massless Dirac fermions in
  graphene}},}\ }\href@noop {} {\bibfield  {journal} {\bibinfo  {journal}
  {Nature}\ }\textbf {\bibinfo {volume} {438}},\ \bibinfo {pages} {197--200}
  (\bibinfo {year} {2005})}\BibitemShut {NoStop}%
\bibitem [{\citenamefont {Zhang}\ \emph {et~al.}(2005)\citenamefont {Zhang},
  \citenamefont {Tan}, \citenamefont {Stormer},\ and\ \citenamefont
  {Kim}}]{Zhang:2005gp}%
  \BibitemOpen
  \bibfield  {author} {\bibinfo {author} {\bibfnamefont {Y.}~\bibnamefont
  {Zhang}}, \bibinfo {author} {\bibfnamefont {Y.}~\bibnamefont {Tan}}, \bibinfo
  {author} {\bibfnamefont {H.}~\bibnamefont {Stormer}}, \ and\ \bibinfo
  {author} {\bibfnamefont {P.}~\bibnamefont {Kim}},\ }\bibfield  {title}
  {\enquote {\bibinfo {title} {{Experimental observation of the quantum Hall
  effect and Berry's phase in graphene}},}\ }\href@noop {} {\bibfield
  {journal} {\bibinfo  {journal} {Nature}\ }\textbf {\bibinfo {volume} {438}},\
  \bibinfo {pages} {201--204} (\bibinfo {year} {2005})}\BibitemShut {NoStop}%
\bibitem [{\citenamefont {Goerbig}(2011)}]{Goerbig:2011un}%
  \BibitemOpen
  \bibfield  {author} {\bibinfo {author} {\bibfnamefont {M.~O.}\ \bibnamefont
  {Goerbig}},\ }\bibfield  {title} {\enquote {\bibinfo {title} {{Electronic
  properties of graphene in a strong magnetic field}},}\ }\href@noop {}
  {\bibfield  {journal} {\bibinfo  {journal} {Reviews Of Modern Physics}\ }
  (\bibinfo {year} {2011})}\BibitemShut {NoStop}%
\bibitem [{\citenamefont {Rickhaus}\ \emph {et~al.}(2015)\citenamefont
  {Rickhaus}, \citenamefont {Liu}, \citenamefont {von~aacute ri}, \citenamefont
  {Weiss}, \citenamefont {Maurand}, \citenamefont {Richter}, \citenamefont
  {nenberger},\ and\ \citenamefont {ter Makk}}]{Rickhaus:2015cp}%
  \BibitemOpen
  \bibfield  {author} {\bibinfo {author} {\bibfnamefont {P.}~\bibnamefont
  {Rickhaus}}, \bibinfo {author} {\bibfnamefont {M.-H.}\ \bibnamefont {Liu}},
  \bibinfo {author} {\bibfnamefont {E.~T.~o.}\ \bibnamefont {von~aacute ri}},
  \bibinfo {author} {\bibfnamefont {M.}~\bibnamefont {Weiss}}, \bibinfo
  {author} {\bibfnamefont {R.}~\bibnamefont {Maurand}}, \bibinfo {author}
  {\bibfnamefont {K.}~\bibnamefont {Richter}}, \bibinfo {author} {\bibfnamefont
  {C.~S.~o.}\ \bibnamefont {nenberger}}, \ and\ \bibinfo {author}
  {\bibfnamefont {P.~e.}\ \bibnamefont {ter Makk}},\ }\bibfield  {title}
  {\enquote {\bibinfo {title} {{Snake trajectories in ultraclean graphene p--n
  junctions}},}\ }\href@noop {} {\bibfield  {journal} {\bibinfo  {journal}
  {Nature Communications}\ }\textbf {\bibinfo {volume} {6}},\ \bibinfo {pages}
  {1--6} (\bibinfo {year} {2015})}\BibitemShut {NoStop}%
\bibitem [{\citenamefont {Chen}\ \emph {et~al.}(2016)\citenamefont {Chen},
  \citenamefont {Han}, \citenamefont {Elahi}, \citenamefont {Habib},
  \citenamefont {Wang}, \citenamefont {Wen}, \citenamefont {Gao}, \citenamefont
  {Taniguchi}, \citenamefont {Watanabe}, \citenamefont {Hone}, \citenamefont
  {Ghosh},\ and\ \citenamefont {Dean}}]{Chen:2016ep}%
  \BibitemOpen
  \bibfield  {author} {\bibinfo {author} {\bibfnamefont {S.}~\bibnamefont
  {Chen}}, \bibinfo {author} {\bibfnamefont {Z.}~\bibnamefont {Han}}, \bibinfo
  {author} {\bibfnamefont {M.~M.}\ \bibnamefont {Elahi}}, \bibinfo {author}
  {\bibfnamefont {K.~M.~M.}\ \bibnamefont {Habib}}, \bibinfo {author}
  {\bibfnamefont {L.}~\bibnamefont {Wang}}, \bibinfo {author} {\bibfnamefont
  {B.}~\bibnamefont {Wen}}, \bibinfo {author} {\bibfnamefont {Y.}~\bibnamefont
  {Gao}}, \bibinfo {author} {\bibfnamefont {T.}~\bibnamefont {Taniguchi}},
  \bibinfo {author} {\bibfnamefont {K.}~\bibnamefont {Watanabe}}, \bibinfo
  {author} {\bibfnamefont {J.}~\bibnamefont {Hone}}, \bibinfo {author}
  {\bibfnamefont {A.~W.}\ \bibnamefont {Ghosh}}, \ and\ \bibinfo {author}
  {\bibfnamefont {C.~R.}\ \bibnamefont {Dean}},\ }\bibfield  {title} {\enquote
  {\bibinfo {title} {{Electron optics with p-n junctions in ballistic
  graphene.}}}\ }\href@noop {} {\bibfield  {journal} {\bibinfo  {journal}
  {Science (New York, NY)}\ }\textbf {\bibinfo {volume} {353}},\ \bibinfo
  {pages} {1522--1525} (\bibinfo {year} {2016})}\BibitemShut {NoStop}%
\bibitem [{\citenamefont {Zhao}\ \emph {et~al.}(2015)\citenamefont {Zhao},
  \citenamefont {Wyrick}, \citenamefont {Natterer}, \citenamefont
  {Rodriguez-Nieva}, \citenamefont {Lewandowski}, \citenamefont {Watanabe},
  \citenamefont {Taniguchi}, \citenamefont {Levitov}, \citenamefont
  {Zhitenev},\ and\ \citenamefont {Stroscio}}]{2015Sci...348..672Z}%
  \BibitemOpen
  \bibfield  {author} {\bibinfo {author} {\bibfnamefont {Y.}~\bibnamefont
  {Zhao}}, \bibinfo {author} {\bibfnamefont {J.}~\bibnamefont {Wyrick}},
  \bibinfo {author} {\bibfnamefont {F.~D.}\ \bibnamefont {Natterer}}, \bibinfo
  {author} {\bibfnamefont {J.~F.}\ \bibnamefont {Rodriguez-Nieva}}, \bibinfo
  {author} {\bibfnamefont {C.}~\bibnamefont {Lewandowski}}, \bibinfo {author}
  {\bibfnamefont {K.}~\bibnamefont {Watanabe}}, \bibinfo {author}
  {\bibfnamefont {T.}~\bibnamefont {Taniguchi}}, \bibinfo {author}
  {\bibfnamefont {L.~S.}\ \bibnamefont {Levitov}}, \bibinfo {author}
  {\bibfnamefont {N.~B.}\ \bibnamefont {Zhitenev}}, \ and\ \bibinfo {author}
  {\bibfnamefont {J.~A.}\ \bibnamefont {Stroscio}},\ }\bibfield  {title}
  {\enquote {\bibinfo {title} {{Creating and probing electron
  whispering-gallery modes in graphene}},}\ }\href@noop {} {\bibfield
  {journal} {\bibinfo  {journal} {Science (New York, NY)}\ }\textbf {\bibinfo
  {volume} {348}},\ \bibinfo {pages} {672--675} (\bibinfo {year}
  {2015})}\BibitemShut {NoStop}%
\bibitem [{\citenamefont {Bandurin}\ \emph {et~al.}(2016)\citenamefont
  {Bandurin}, \citenamefont {Torre}, \citenamefont {Kumar}, \citenamefont
  {Ben~Shalom}, \citenamefont {Tomadin}, \citenamefont {Principi},
  \citenamefont {Auton}, \citenamefont {Khestanova}, \citenamefont {Novoselov},
  \citenamefont {Grigorieva}, \citenamefont {Ponomarenko}, \citenamefont
  {Geim},\ and\ \citenamefont {Polini}}]{Bandurin:2016cp}%
  \BibitemOpen
  \bibfield  {author} {\bibinfo {author} {\bibfnamefont {D.~A.}\ \bibnamefont
  {Bandurin}}, \bibinfo {author} {\bibfnamefont {I.}~\bibnamefont {Torre}},
  \bibinfo {author} {\bibfnamefont {R.~K.}\ \bibnamefont {Kumar}}, \bibinfo
  {author} {\bibfnamefont {M.}~\bibnamefont {Ben~Shalom}}, \bibinfo {author}
  {\bibfnamefont {A.}~\bibnamefont {Tomadin}}, \bibinfo {author} {\bibfnamefont
  {A.}~\bibnamefont {Principi}}, \bibinfo {author} {\bibfnamefont {G.~H.}\
  \bibnamefont {Auton}}, \bibinfo {author} {\bibfnamefont {E.}~\bibnamefont
  {Khestanova}}, \bibinfo {author} {\bibfnamefont {K.~S.}\ \bibnamefont
  {Novoselov}}, \bibinfo {author} {\bibfnamefont {I.~V.}\ \bibnamefont
  {Grigorieva}}, \bibinfo {author} {\bibfnamefont {L.~A.}\ \bibnamefont
  {Ponomarenko}}, \bibinfo {author} {\bibfnamefont {A.~K.}\ \bibnamefont
  {Geim}}, \ and\ \bibinfo {author} {\bibfnamefont {M.}~\bibnamefont
  {Polini}},\ }\bibfield  {title} {\enquote {\bibinfo {title} {{Negative local
  resistance caused by viscous electron backflow in graphene}},}\ }\href@noop
  {} {\bibfield  {journal} {\bibinfo  {journal} {Science (New York, NY)}\
  }\textbf {\bibinfo {volume} {351}},\ \bibinfo {pages} {1055--1058} (\bibinfo
  {year} {2016})}\BibitemShut {NoStop}%
\bibitem [{\citenamefont {Crossno}\ \emph {et~al.}(2016)\citenamefont
  {Crossno}, \citenamefont {Shi}, \citenamefont {Wang}, \citenamefont {Liu},
  \citenamefont {Harzheim}, \citenamefont {Lucas}, \citenamefont {Sachdev},
  \citenamefont {Kim}, \citenamefont {Taniguchi}, \citenamefont {Watanabe},
  \citenamefont {Ohki},\ and\ \citenamefont {Fong}}]{Crossno:2016iy}%
  \BibitemOpen
  \bibfield  {author} {\bibinfo {author} {\bibfnamefont {J.}~\bibnamefont
  {Crossno}}, \bibinfo {author} {\bibfnamefont {J.~K.}\ \bibnamefont {Shi}},
  \bibinfo {author} {\bibfnamefont {K.}~\bibnamefont {Wang}}, \bibinfo {author}
  {\bibfnamefont {X.}~\bibnamefont {Liu}}, \bibinfo {author} {\bibfnamefont
  {A.}~\bibnamefont {Harzheim}}, \bibinfo {author} {\bibfnamefont
  {A.}~\bibnamefont {Lucas}}, \bibinfo {author} {\bibfnamefont
  {S.}~\bibnamefont {Sachdev}}, \bibinfo {author} {\bibfnamefont
  {P.}~\bibnamefont {Kim}}, \bibinfo {author} {\bibfnamefont {T.}~\bibnamefont
  {Taniguchi}}, \bibinfo {author} {\bibfnamefont {K.}~\bibnamefont {Watanabe}},
  \bibinfo {author} {\bibfnamefont {T.~A.}\ \bibnamefont {Ohki}}, \ and\
  \bibinfo {author} {\bibfnamefont {K.~C.}\ \bibnamefont {Fong}},\ }\bibfield
  {title} {\enquote {\bibinfo {title} {{Observation of the Dirac fluid and the
  breakdown of the Wiedemann-Franz law in graphene}},}\ }\href@noop {}
  {\bibfield  {journal} {\bibinfo  {journal} {Science (New York, NY)}\ }\textbf
  {\bibinfo {volume} {351}},\ \bibinfo {pages} {1058--1061} (\bibinfo {year}
  {2016})}\BibitemShut {NoStop}%
\bibitem [{\citenamefont {Ghahari}\ \emph {et~al.}(2016)\citenamefont
  {Ghahari}, \citenamefont {Xie}, \citenamefont {Taniguchi}, \citenamefont
  {Watanabe}, \citenamefont {Foster},\ and\ \citenamefont
  {Kim}}]{Ghahari:2016df}%
  \BibitemOpen
  \bibfield  {author} {\bibinfo {author} {\bibfnamefont {F.}~\bibnamefont
  {Ghahari}}, \bibinfo {author} {\bibfnamefont {H.-Y.}\ \bibnamefont {Xie}},
  \bibinfo {author} {\bibfnamefont {T.}~\bibnamefont {Taniguchi}}, \bibinfo
  {author} {\bibfnamefont {K.}~\bibnamefont {Watanabe}}, \bibinfo {author}
  {\bibfnamefont {M.~S.}\ \bibnamefont {Foster}}, \ and\ \bibinfo {author}
  {\bibfnamefont {P.}~\bibnamefont {Kim}},\ }\bibfield  {title} {\enquote
  {\bibinfo {title} {{Enhanced Thermoelectric Power in Graphene: Violation of
  the Mott Relation by Inelastic Scattering}},}\ }\href@noop {} {\bibfield
  {journal} {\bibinfo  {journal} {Physical Review Letters}\ }\textbf {\bibinfo
  {volume} {116}},\ \bibinfo {pages} {136802} (\bibinfo {year}
  {2016})}\BibitemShut {NoStop}%
\bibitem [{\citenamefont {Schwierz}(2010)}]{Schwierz:2010ix}%
  \BibitemOpen
  \bibfield  {author} {\bibinfo {author} {\bibfnamefont {F.}~\bibnamefont
  {Schwierz}},\ }\bibfield  {title} {\enquote {\bibinfo {title} {{Graphene
  transistors}},}\ }\href@noop {} {\bibfield  {journal} {\bibinfo  {journal}
  {Nature Nanotechnology}\ }\textbf {\bibinfo {volume} {5}},\ \bibinfo {pages}
  {487--496} (\bibinfo {year} {2010})}\BibitemShut {NoStop}%
\bibitem [{\citenamefont {Palacios}\ \emph {et~al.}(2010)\citenamefont
  {Palacios}, \citenamefont {Hsu},\ and\ \citenamefont
  {Wang}}]{Palacios:2010dw}%
  \BibitemOpen
  \bibfield  {author} {\bibinfo {author} {\bibfnamefont {T.}~\bibnamefont
  {Palacios}}, \bibinfo {author} {\bibfnamefont {A.}~\bibnamefont {Hsu}}, \
  and\ \bibinfo {author} {\bibfnamefont {H.}~\bibnamefont {Wang}},\ }\bibfield
  {title} {\enquote {\bibinfo {title} {{Applications of Graphene Devices in RF
  Communications}},}\ }\href@noop {} {\bibfield  {journal} {\bibinfo  {journal}
  {Ieee Communications Magazine}\ }\textbf {\bibinfo {volume} {48}},\ \bibinfo
  {pages} {122--128} (\bibinfo {year} {2010})}\BibitemShut {NoStop}%
\bibitem [{\citenamefont {Glazov}\ and\ \citenamefont
  {Ganichev}(2014)}]{Glazov:2014cp}%
  \BibitemOpen
  \bibfield  {author} {\bibinfo {author} {\bibfnamefont {M.~M.}\ \bibnamefont
  {Glazov}}\ and\ \bibinfo {author} {\bibfnamefont {S.~D.}\ \bibnamefont
  {Ganichev}},\ }\bibfield  {title} {\enquote {\bibinfo {title} {{Physics
  Reports}},}\ }\href@noop {} {\bibfield  {journal} {\bibinfo  {journal}
  {Physics Reports-Review Section Of Physics Letters}\ }\textbf {\bibinfo
  {volume} {535}},\ \bibinfo {pages} {101--138} (\bibinfo {year}
  {2014})}\BibitemShut {NoStop}%
\bibitem [{\citenamefont {Otsuji}\ \emph {et~al.}(2012)\citenamefont {Otsuji},
  \citenamefont {Boubanga~Tombet}, \citenamefont {Satou}, \citenamefont
  {Fukidome}, \citenamefont {Suemitsu}, \citenamefont {Sano}, \citenamefont
  {Popov}, \citenamefont {Ryzhii},\ and\ \citenamefont
  {Ryzhii}}]{Otsuji:2012hn}%
  \BibitemOpen
  \bibfield  {author} {\bibinfo {author} {\bibfnamefont {T.}~\bibnamefont
  {Otsuji}}, \bibinfo {author} {\bibfnamefont {S.~A.}\ \bibnamefont
  {Boubanga~Tombet}}, \bibinfo {author} {\bibfnamefont {A.}~\bibnamefont
  {Satou}}, \bibinfo {author} {\bibfnamefont {H.}~\bibnamefont {Fukidome}},
  \bibinfo {author} {\bibfnamefont {M.}~\bibnamefont {Suemitsu}}, \bibinfo
  {author} {\bibfnamefont {E.}~\bibnamefont {Sano}}, \bibinfo {author}
  {\bibfnamefont {V.}~\bibnamefont {Popov}}, \bibinfo {author} {\bibfnamefont
  {M.}~\bibnamefont {Ryzhii}}, \ and\ \bibinfo {author} {\bibfnamefont
  {V.}~\bibnamefont {Ryzhii}},\ }\bibfield  {title} {\enquote {\bibinfo {title}
  {{Graphene materials and devices in terahertz science and technology}},}\
  }\href@noop {} {\bibfield  {journal} {\bibinfo  {journal} {MRS Bulletin}\
  }\textbf {\bibinfo {volume} {37}},\ \bibinfo {pages} {1235--1243} (\bibinfo
  {year} {2012})}\BibitemShut {NoStop}%
\bibitem [{\citenamefont {Koppens}\ \emph {et~al.}(2014)\citenamefont
  {Koppens}, \citenamefont {Mueller}, \citenamefont {Avouris}, \citenamefont
  {Ferrari}, \citenamefont {Vitiello},\ and\ \citenamefont
  {Polini}}]{Koppens:2014dy}%
  \BibitemOpen
  \bibfield  {author} {\bibinfo {author} {\bibfnamefont {F.~H.~L.}\
  \bibnamefont {Koppens}}, \bibinfo {author} {\bibfnamefont {T.}~\bibnamefont
  {Mueller}}, \bibinfo {author} {\bibfnamefont {P.}~\bibnamefont {Avouris}},
  \bibinfo {author} {\bibfnamefont {A.~C.}\ \bibnamefont {Ferrari}}, \bibinfo
  {author} {\bibfnamefont {M.~S.}\ \bibnamefont {Vitiello}}, \ and\ \bibinfo
  {author} {\bibfnamefont {M.}~\bibnamefont {Polini}},\ }\bibfield  {title}
  {\enquote {\bibinfo {title} {{Photodetectors based on graphene,
  othertwo-dimensional materials and hybrid systems}},}\ }\href@noop {}
  {\bibfield  {journal} {\bibinfo  {journal} {Nature Nanotechnology}\ }\textbf
  {\bibinfo {volume} {9}},\ \bibinfo {pages} {780--793} (\bibinfo {year}
  {2014})}\BibitemShut {NoStop}%
\bibitem [{\citenamefont {Cheng}\ \emph {et~al.}(2012)\citenamefont {Cheng},
  \citenamefont {Bai}, \citenamefont {Liao}, \citenamefont {Zhou},
  \citenamefont {Chen}, \citenamefont {Liu}, \citenamefont {Lin}, \citenamefont
  {Jiang}, \citenamefont {Huang},\ and\ \citenamefont
  {Duan}}]{2012PNAS..10911588C}%
  \BibitemOpen
  \bibfield  {author} {\bibinfo {author} {\bibfnamefont {R.}~\bibnamefont
  {Cheng}}, \bibinfo {author} {\bibfnamefont {J.}~\bibnamefont {Bai}}, \bibinfo
  {author} {\bibfnamefont {L.}~\bibnamefont {Liao}}, \bibinfo {author}
  {\bibfnamefont {H.}~\bibnamefont {Zhou}}, \bibinfo {author} {\bibfnamefont
  {Y.}~\bibnamefont {Chen}}, \bibinfo {author} {\bibfnamefont {L.}~\bibnamefont
  {Liu}}, \bibinfo {author} {\bibfnamefont {Y.-C.}\ \bibnamefont {Lin}},
  \bibinfo {author} {\bibfnamefont {S.}~\bibnamefont {Jiang}}, \bibinfo
  {author} {\bibfnamefont {Y.}~\bibnamefont {Huang}}, \ and\ \bibinfo {author}
  {\bibfnamefont {X.}~\bibnamefont {Duan}},\ }\bibfield  {title} {\enquote
  {\bibinfo {title} {{High-frequency self-aligned graphene transistors with
  transferred gate stacks}},}\ }\href@noop {} {\bibfield  {journal} {\bibinfo
  {journal} {Proceedings Of The National Academy Of Sciences Of The United
  States Of America}\ }\textbf {\bibinfo {volume} {109}},\ \bibinfo {pages}
  {11588--11592} (\bibinfo {year} {2012})}\BibitemShut {NoStop}%
\bibitem [{\citenamefont {Habibpour}\ \emph {et~al.}()\citenamefont
  {Habibpour}, \citenamefont {Cherednichenko}, \citenamefont {Vukusic},
  \citenamefont {Yhland},\ and\ \citenamefont {Stake}}]{Habibpour:fz}%
  \BibitemOpen
  \bibfield  {author} {\bibinfo {author} {\bibfnamefont {O.}~\bibnamefont
  {Habibpour}}, \bibinfo {author} {\bibfnamefont {S.}~\bibnamefont
  {Cherednichenko}}, \bibinfo {author} {\bibfnamefont {J.}~\bibnamefont
  {Vukusic}}, \bibinfo {author} {\bibfnamefont {K.}~\bibnamefont {Yhland}}, \
  and\ \bibinfo {author} {\bibfnamefont {J.}~\bibnamefont {Stake}},\ }\bibfield
   {title} {\enquote {\bibinfo {title} {{A Subharmonic Graphene FET Mixer}},}\
  }\href@noop {} {\bibfield  {journal} {\bibinfo  {journal} {IEEE Electron
  Device Letters}\ }\textbf {\bibinfo {volume} {33}},\ \bibinfo {pages}
  {71--73}}\BibitemShut {NoStop}%
\bibitem [{\citenamefont {Wang}\ \emph {et~al.}(2009)\citenamefont {Wang},
  \citenamefont {Nezich}, \citenamefont {Kong},\ and\ \citenamefont
  {Palacios}}]{Wang:2009jq}%
  \BibitemOpen
  \bibfield  {author} {\bibinfo {author} {\bibfnamefont {H.}~\bibnamefont
  {Wang}}, \bibinfo {author} {\bibfnamefont {D.}~\bibnamefont {Nezich}},
  \bibinfo {author} {\bibfnamefont {J.}~\bibnamefont {Kong}}, \ and\ \bibinfo
  {author} {\bibfnamefont {T.}~\bibnamefont {Palacios}},\ }\bibfield  {title}
  {\enquote {\bibinfo {title} {{Graphene Frequency Multipliers}},}\ }\href@noop
  {} {\bibfield  {journal} {\bibinfo  {journal} {IEEE Electron Device Letters}\
  }\textbf {\bibinfo {volume} {30}},\ \bibinfo {pages} {547--549} (\bibinfo
  {year} {2009})}\BibitemShut {NoStop}%
\bibitem [{\citenamefont {Vicarelli}\ \emph {et~al.}(2012)\citenamefont
  {Vicarelli}, \citenamefont {Vitiello}, \citenamefont {Coquillat},
  \citenamefont {Lombardo}, \citenamefont {Ferrari}, \citenamefont {Knap},
  \citenamefont {Polini}, \citenamefont {Pellegrini},\ and\ \citenamefont
  {Tredicucci}}]{Vicarelli:2012ch}%
  \BibitemOpen
  \bibfield  {author} {\bibinfo {author} {\bibfnamefont {L.}~\bibnamefont
  {Vicarelli}}, \bibinfo {author} {\bibfnamefont {M.~S.}\ \bibnamefont
  {Vitiello}}, \bibinfo {author} {\bibfnamefont {D.}~\bibnamefont {Coquillat}},
  \bibinfo {author} {\bibfnamefont {A.}~\bibnamefont {Lombardo}}, \bibinfo
  {author} {\bibfnamefont {A.~C.}\ \bibnamefont {Ferrari}}, \bibinfo {author}
  {\bibfnamefont {W.}~\bibnamefont {Knap}}, \bibinfo {author} {\bibfnamefont
  {M.}~\bibnamefont {Polini}}, \bibinfo {author} {\bibfnamefont
  {V.}~\bibnamefont {Pellegrini}}, \ and\ \bibinfo {author} {\bibfnamefont
  {A.}~\bibnamefont {Tredicucci}},\ }\bibfield  {title} {\enquote {\bibinfo
  {title} {{Graphene field-effect transistors as room-temperature terahertz
  detectors}},}\ }\href@noop {} {\bibfield  {journal} {\bibinfo  {journal}
  {Nature Materials}\ }\textbf {\bibinfo {volume} {11}},\ \bibinfo {pages}
  {865--871} (\bibinfo {year} {2012})}\BibitemShut {NoStop}%
\bibitem [{\citenamefont {Mittendorff}\ \emph {et~al.}(2013)\citenamefont
  {Mittendorff}, \citenamefont {Winnerl}, \citenamefont {Kamann}, \citenamefont
  {Eroms}, \citenamefont {Weiss}, \citenamefont {Schneider},\ and\
  \citenamefont {Helm}}]{Mittendorff:2013gk}%
  \BibitemOpen
  \bibfield  {author} {\bibinfo {author} {\bibfnamefont {M.}~\bibnamefont
  {Mittendorff}}, \bibinfo {author} {\bibfnamefont {S.}~\bibnamefont
  {Winnerl}}, \bibinfo {author} {\bibfnamefont {J.}~\bibnamefont {Kamann}},
  \bibinfo {author} {\bibfnamefont {J.}~\bibnamefont {Eroms}}, \bibinfo
  {author} {\bibfnamefont {D.}~\bibnamefont {Weiss}}, \bibinfo {author}
  {\bibfnamefont {H.}~\bibnamefont {Schneider}}, \ and\ \bibinfo {author}
  {\bibfnamefont {M.}~\bibnamefont {Helm}},\ }\bibfield  {title} {\enquote
  {\bibinfo {title} {{Ultrafast graphene-based broadband THz detector}},}\
  }\href@noop {} {\bibfield  {journal} {\bibinfo  {journal} {Applied Physics
  Letters}\ }\textbf {\bibinfo {volume} {103}},\ \bibinfo {pages} {021113}
  (\bibinfo {year} {2013})}\BibitemShut {NoStop}%
\bibitem [{\citenamefont {Cai}\ \emph {et~al.}(2014)\citenamefont {Cai},
  \citenamefont {Sushkov}, \citenamefont {Suess}, \citenamefont {Jadidi},
  \citenamefont {Jenkins}, \citenamefont {Nyakiti}, \citenamefont {Myers-Ward},
  \citenamefont {Li}, \citenamefont {Yan}, \citenamefont {Gaskill},
  \citenamefont {Murphy}, \citenamefont {Drew},\ and\ \citenamefont
  {Fuhrer}}]{Cai:2014hh}%
  \BibitemOpen
  \bibfield  {author} {\bibinfo {author} {\bibfnamefont {X.}~\bibnamefont
  {Cai}}, \bibinfo {author} {\bibfnamefont {A.~B.}\ \bibnamefont {Sushkov}},
  \bibinfo {author} {\bibfnamefont {R.~J.}\ \bibnamefont {Suess}}, \bibinfo
  {author} {\bibfnamefont {M.~M.}\ \bibnamefont {Jadidi}}, \bibinfo {author}
  {\bibfnamefont {G.~S.}\ \bibnamefont {Jenkins}}, \bibinfo {author}
  {\bibfnamefont {L.~O.}\ \bibnamefont {Nyakiti}}, \bibinfo {author}
  {\bibfnamefont {R.~L.}\ \bibnamefont {Myers-Ward}}, \bibinfo {author}
  {\bibfnamefont {S.}~\bibnamefont {Li}}, \bibinfo {author} {\bibfnamefont
  {J.}~\bibnamefont {Yan}}, \bibinfo {author} {\bibfnamefont {D.~K.}\
  \bibnamefont {Gaskill}}, \bibinfo {author} {\bibfnamefont {T.~E.}\
  \bibnamefont {Murphy}}, \bibinfo {author} {\bibfnamefont {H.~D.}\
  \bibnamefont {Drew}}, \ and\ \bibinfo {author} {\bibfnamefont {M.~S.}\
  \bibnamefont {Fuhrer}},\ }\bibfield  {title} {\enquote {\bibinfo {title}
  {{Sensitive room-temperature terahertz detectionvia the photothermoelectric
  effect in graphene}},}\ }\href@noop {} {\bibfield  {journal} {\bibinfo
  {journal} {Nature Nanotechnology}\ }\textbf {\bibinfo {volume} {9}},\
  \bibinfo {pages} {814 --819} (\bibinfo {year} {2014})}\BibitemShut {NoStop}%
\bibitem [{\citenamefont {Zak}\ \emph {et~al.}(2014)\citenamefont {Zak},
  \citenamefont {Andersson}, \citenamefont {Bauer}, \citenamefont {Matukas},
  \citenamefont {Lisauskas}, \citenamefont {Roskos},\ and\ \citenamefont
  {Stake}}]{Zak:2014gc}%
  \BibitemOpen
  \bibfield  {author} {\bibinfo {author} {\bibfnamefont {A.}~\bibnamefont
  {Zak}}, \bibinfo {author} {\bibfnamefont {M.~A.}\ \bibnamefont {Andersson}},
  \bibinfo {author} {\bibfnamefont {M.}~\bibnamefont {Bauer}}, \bibinfo
  {author} {\bibfnamefont {J.}~\bibnamefont {Matukas}}, \bibinfo {author}
  {\bibfnamefont {A.}~\bibnamefont {Lisauskas}}, \bibinfo {author}
  {\bibfnamefont {H.~G.}\ \bibnamefont {Roskos}}, \ and\ \bibinfo {author}
  {\bibfnamefont {J.}~\bibnamefont {Stake}},\ }\bibfield  {title} {\enquote
  {\bibinfo {title} {{Antenna-Integrated 0.6 THz FET Direct Detectors Based on
  CVD Graphene}},}\ }\href@noop {} {\bibfield  {journal} {\bibinfo  {journal}
  {Nano Letters}\ }\textbf {\bibinfo {volume} {14}},\ \bibinfo {pages}
  {5834--5838} (\bibinfo {year} {2014})}\BibitemShut {NoStop}%
\bibitem [{\citenamefont {Prada}\ \emph {et~al.}(2009)\citenamefont {Prada},
  \citenamefont {San-Jose},\ and\ \citenamefont {Schomerus}}]{Prada:2009jo}%
  \BibitemOpen
  \bibfield  {author} {\bibinfo {author} {\bibfnamefont {E.}~\bibnamefont
  {Prada}}, \bibinfo {author} {\bibfnamefont {P.}~\bibnamefont {San-Jose}}, \
  and\ \bibinfo {author} {\bibfnamefont {H.}~\bibnamefont {Schomerus}},\
  }\bibfield  {title} {\enquote {\bibinfo {title} {{Quantum pumping in
  graphene}},}\ }\href@noop {} {\bibfield  {journal} {\bibinfo  {journal}
  {Physical Review B}\ }\textbf {\bibinfo {volume} {80}},\ \bibinfo {pages}
  {245414} (\bibinfo {year} {2009})}\BibitemShut {NoStop}%
\bibitem [{\citenamefont {Foa~Torres}\ \emph {et~al.}(2011)\citenamefont
  {Foa~Torres}, \citenamefont {Calvo}, \citenamefont {Rocha},\ and\
  \citenamefont {Cuniberti}}]{FoaTorres:2011kf}%
  \BibitemOpen
  \bibfield  {author} {\bibinfo {author} {\bibfnamefont {L.~E.~F.}\
  \bibnamefont {Foa~Torres}}, \bibinfo {author} {\bibfnamefont {H.~L.}\
  \bibnamefont {Calvo}}, \bibinfo {author} {\bibfnamefont {C.~G.}\ \bibnamefont
  {Rocha}}, \ and\ \bibinfo {author} {\bibfnamefont {G.}~\bibnamefont
  {Cuniberti}},\ }\bibfield  {title} {\enquote {\bibinfo {title} {{Enhancing
  single-parameter quantum charge pumping in carbon-based devices}},}\
  }\href@noop {} {\bibfield  {journal} {\bibinfo  {journal} {Applied Physics
  Letters}\ }\textbf {\bibinfo {volume} {99}},\ \bibinfo {pages} {092102}
  (\bibinfo {year} {2011})}\BibitemShut {NoStop}%
\bibitem [{\citenamefont {San-Jose}\ \emph {et~al.}(2011)\citenamefont
  {San-Jose}, \citenamefont {Prada}, \citenamefont {Kohler},\ and\
  \citenamefont {Schomerus}}]{SanJose:2011fe}%
  \BibitemOpen
  \bibfield  {author} {\bibinfo {author} {\bibfnamefont {P.}~\bibnamefont
  {San-Jose}}, \bibinfo {author} {\bibfnamefont {E.}~\bibnamefont {Prada}},
  \bibinfo {author} {\bibfnamefont {S.}~\bibnamefont {Kohler}}, \ and\ \bibinfo
  {author} {\bibfnamefont {H.}~\bibnamefont {Schomerus}},\ }\bibfield  {title}
  {\enquote {\bibinfo {title} {{Single-parameter pumping in graphene}},}\
  }\href@noop {} {\bibfield  {journal} {\bibinfo  {journal} {Physical Review
  B}\ }\textbf {\bibinfo {volume} {84}},\ \bibinfo {pages} {155408} (\bibinfo
  {year} {2011})}\BibitemShut {NoStop}%
\bibitem [{\citenamefont {Zhu}\ and\ \citenamefont {Lai}(2011)}]{Zhu:2011cr}%
  \BibitemOpen
  \bibfield  {author} {\bibinfo {author} {\bibfnamefont {R.}~\bibnamefont
  {Zhu}}\ and\ \bibinfo {author} {\bibfnamefont {M.}~\bibnamefont {Lai}},\
  }\bibfield  {title} {\enquote {\bibinfo {title} {{Pumped shot noise in
  adiabatically modulated graphene-based double-barrier structures}},}\
  }\href@noop {} {\bibfield  {journal} {\bibinfo  {journal} {Journal Of
  Physics-Condensed Matter}\ }\textbf {\bibinfo {volume} {23}},\ \bibinfo
  {pages} {455302} (\bibinfo {year} {2011})}\BibitemShut {NoStop}%
\bibitem [{\citenamefont {San-Jose}\ \emph {et~al.}(2012)\citenamefont
  {San-Jose}, \citenamefont {Prada}, \citenamefont {Schomerus},\ and\
  \citenamefont {Kohler}}]{SanJose:2012bw}%
  \BibitemOpen
  \bibfield  {author} {\bibinfo {author} {\bibfnamefont {P.}~\bibnamefont
  {San-Jose}}, \bibinfo {author} {\bibfnamefont {E.}~\bibnamefont {Prada}},
  \bibinfo {author} {\bibfnamefont {H.}~\bibnamefont {Schomerus}}, \ and\
  \bibinfo {author} {\bibfnamefont {S.}~\bibnamefont {Kohler}},\ }\bibfield
  {title} {\enquote {\bibinfo {title} {{Laser-induced quantum pumping in
  graphene}},}\ }\href@noop {} {\bibfield  {journal} {\bibinfo  {journal}
  {Applied Physics Letters}\ }\textbf {\bibinfo {volume} {101}},\ \bibinfo
  {pages} {153506} (\bibinfo {year} {2012})}\BibitemShut {NoStop}%
\bibitem [{\citenamefont {Mikhailov}\ and\ \citenamefont
  {Ziegler}(2007)}]{Mikhailov:2007ft}%
  \BibitemOpen
  \bibfield  {author} {\bibinfo {author} {\bibfnamefont {S.~A.}\ \bibnamefont
  {Mikhailov}}\ and\ \bibinfo {author} {\bibfnamefont {K.}~\bibnamefont
  {Ziegler}},\ }\bibfield  {title} {\enquote {\bibinfo {title} {{New
  Electromagnetic Mode in Graphene}},}\ }\href@noop {} {\bibfield  {journal}
  {\bibinfo  {journal} {Physical Review Letters}\ }\textbf {\bibinfo {volume}
  {99}},\ \bibinfo {pages} {016803} (\bibinfo {year} {2007})}\BibitemShut
  {NoStop}%
\bibitem [{\citenamefont {Mikhailov}\ and\ \citenamefont
  {Ziegler}(2008)}]{Mikhailov:2008ig}%
  \BibitemOpen
  \bibfield  {author} {\bibinfo {author} {\bibfnamefont {S.~A.}\ \bibnamefont
  {Mikhailov}}\ and\ \bibinfo {author} {\bibfnamefont {K.}~\bibnamefont
  {Ziegler}},\ }\bibfield  {title} {\enquote {\bibinfo {title} {{Nonlinear
  electromagnetic response of graphene: frequency multiplication and the
  self-consistent-field effects}},}\ }\href@noop {} {\bibfield  {journal}
  {\bibinfo  {journal} {Journal Of Physics-Condensed Matter}\ }\textbf
  {\bibinfo {volume} {20}},\ \bibinfo {pages} {384204} (\bibinfo {year}
  {2008})}\BibitemShut {NoStop}%
\bibitem [{\citenamefont {Syzranov}\ \emph {et~al.}(2008)\citenamefont
  {Syzranov}, \citenamefont {Fistul},\ and\ \citenamefont
  {Efetov}}]{Syzranov:2008dg}%
  \BibitemOpen
  \bibfield  {author} {\bibinfo {author} {\bibfnamefont {S.~V.}\ \bibnamefont
  {Syzranov}}, \bibinfo {author} {\bibfnamefont {M.~V.}\ \bibnamefont
  {Fistul}}, \ and\ \bibinfo {author} {\bibfnamefont {K.~B.}\ \bibnamefont
  {Efetov}},\ }\bibfield  {title} {\enquote {\bibinfo {title} {{Effect of
  radiation on transport in graphene}},}\ }\href@noop {} {\bibfield  {journal}
  {\bibinfo  {journal} {Physical Review B}\ }\textbf {\bibinfo {volume} {78}},\
  \bibinfo {pages} {045407} (\bibinfo {year} {2008})}\BibitemShut {NoStop}%
\bibitem [{\citenamefont {Calvo}\ \emph {et~al.}(2012)\citenamefont {Calvo},
  \citenamefont {Perez-Piskunow}, \citenamefont {Roche},\ and\ \citenamefont
  {Foa~Torres}}]{Calvo:2012bg}%
  \BibitemOpen
  \bibfield  {author} {\bibinfo {author} {\bibfnamefont {H.~L.}\ \bibnamefont
  {Calvo}}, \bibinfo {author} {\bibfnamefont {P.~M.}\ \bibnamefont
  {Perez-Piskunow}}, \bibinfo {author} {\bibfnamefont {S.}~\bibnamefont
  {Roche}}, \ and\ \bibinfo {author} {\bibfnamefont {L.~E.~F.}\ \bibnamefont
  {Foa~Torres}},\ }\bibfield  {title} {\enquote {\bibinfo {title}
  {{Laser-induced effects on the electronic features of graphene
  nanoribbons}},}\ }\href@noop {} {\bibfield  {journal} {\bibinfo  {journal}
  {Applied Physics Letters}\ }\textbf {\bibinfo {volume} {101}},\ \bibinfo
  {pages} {253506} (\bibinfo {year} {2012})}\BibitemShut {NoStop}%
\bibitem [{\citenamefont {Al-Naib}\ \emph {et~al.}(2014)\citenamefont
  {Al-Naib}, \citenamefont {Sipe},\ and\ \citenamefont
  {Dignam}}]{AlNaib:2014jj}%
  \BibitemOpen
  \bibfield  {author} {\bibinfo {author} {\bibfnamefont {I.}~\bibnamefont
  {Al-Naib}}, \bibinfo {author} {\bibfnamefont {J.~E.}\ \bibnamefont {Sipe}}, \
  and\ \bibinfo {author} {\bibfnamefont {M.~M.}\ \bibnamefont {Dignam}},\
  }\bibfield  {title} {\enquote {\bibinfo {title} {{High harmonic generation in
  undoped graphene: Interplay of inter- and intraband dynamics}},}\ }\href@noop
  {} {\bibfield  {journal} {\bibinfo  {journal} {Physical Review B}\ }\textbf
  {\bibinfo {volume} {90}},\ \bibinfo {pages} {245423} (\bibinfo {year}
  {2014})}\BibitemShut {NoStop}%
\bibitem [{\citenamefont {Sinha}\ and\ \citenamefont
  {Biswas}(2012)}]{Sinha:2012fx}%
  \BibitemOpen
  \bibfield  {author} {\bibinfo {author} {\bibfnamefont {C.}~\bibnamefont
  {Sinha}}\ and\ \bibinfo {author} {\bibfnamefont {R.}~\bibnamefont {Biswas}},\
  }\bibfield  {title} {\enquote {\bibinfo {title} {{Transmission of electron
  through monolayer graphene laser barrier}},}\ }\href@noop {} {\bibfield
  {journal} {\bibinfo  {journal} {Applied Physics Letters}\ }\textbf {\bibinfo
  {volume} {100}},\ \bibinfo {pages} {183107} (\bibinfo {year}
  {2012})}\BibitemShut {NoStop}%
\bibitem [{\citenamefont {Trauzettel}\ \emph {et~al.}(2007)\citenamefont
  {Trauzettel}, \citenamefont {Blanter},\ and\ \citenamefont
  {Morpurgo}}]{Trauzettel:2007kq}%
  \BibitemOpen
  \bibfield  {author} {\bibinfo {author} {\bibfnamefont {B.}~\bibnamefont
  {Trauzettel}}, \bibinfo {author} {\bibfnamefont {Y.~M.}\ \bibnamefont
  {Blanter}}, \ and\ \bibinfo {author} {\bibfnamefont {A.~F.}\ \bibnamefont
  {Morpurgo}},\ }\bibfield  {title} {\enquote {\bibinfo {title}
  {{Photon-assisted electron transport in graphene: Scattering theory
  analysis}},}\ }\href@noop {} {\bibfield  {journal} {\bibinfo  {journal}
  {Physical Review B}\ }\textbf {\bibinfo {volume} {75}},\ \bibinfo {pages}
  {035305} (\bibinfo {year} {2007})}\BibitemShut {NoStop}%
\bibitem [{\citenamefont {Zeb}\ and\ \citenamefont {Tahir}(2008)}]{Zeb:2008ka}%
  \BibitemOpen
  \bibfield  {author} {\bibinfo {author} {\bibfnamefont {M.~A.}\ \bibnamefont
  {Zeb}}\ and\ \bibinfo {author} {\bibfnamefont {M.}~\bibnamefont {Tahir}},\
  }\bibfield  {title} {\enquote {\bibinfo {title} {{Chiral tunneling through a
  time-periodic potential in monolayer graphene}},}\ }\href@noop {} {\bibfield
  {journal} {\bibinfo  {journal} {Physical Review B}\ }\textbf {\bibinfo
  {volume} {78}},\ \bibinfo {pages} {1--7} (\bibinfo {year}
  {2008})}\BibitemShut {NoStop}%
\bibitem [{\citenamefont {Rocha}\ \emph {et~al.}(2010)\citenamefont {Rocha},
  \citenamefont {Torres},\ and\ \citenamefont {Cuniberti}}]{Rocha:2010ho}%
  \BibitemOpen
  \bibfield  {author} {\bibinfo {author} {\bibfnamefont {C.~G.}\ \bibnamefont
  {Rocha}}, \bibinfo {author} {\bibfnamefont {L.~E. F.~F.}\ \bibnamefont
  {Torres}}, \ and\ \bibinfo {author} {\bibfnamefont {G.}~\bibnamefont
  {Cuniberti}},\ }\bibfield  {title} {\enquote {\bibinfo {title} {{ac transport
  in graphene-based Fabry-P{\'e}rot devices}},}\ }\href@noop {} {\bibfield
  {journal} {\bibinfo  {journal} {Physical Review B}\ }\textbf {\bibinfo
  {volume} {81}},\ \bibinfo {pages} {115435} (\bibinfo {year}
  {2010})}\BibitemShut {NoStop}%
\bibitem [{\citenamefont {Savel'ev}\ \emph {et~al.}(2012)\citenamefont
  {Savel'ev}, \citenamefont {H{\"a}usler},\ and\ \citenamefont
  {H{\"a}nggi}}]{Savelev:2012dg}%
  \BibitemOpen
  \bibfield  {author} {\bibinfo {author} {\bibfnamefont {S.~E.}\ \bibnamefont
  {Savel'ev}}, \bibinfo {author} {\bibfnamefont {W.}~\bibnamefont
  {H{\"a}usler}}, \ and\ \bibinfo {author} {\bibfnamefont {P.}~\bibnamefont
  {H{\"a}nggi}},\ }\bibfield  {title} {\enquote {\bibinfo {title} {{Current
  Resonances in Graphene with Time-Dependent Potential Barriers}},}\
  }\href@noop {} {\bibfield  {journal} {\bibinfo  {journal} {Physical Review
  Letters}\ }\textbf {\bibinfo {volume} {109}},\ \bibinfo {pages} {226602}
  (\bibinfo {year} {2012})}\BibitemShut {NoStop}%
\bibitem [{\citenamefont {Lu}\ \emph {et~al.}(2012)\citenamefont {Lu},
  \citenamefont {Wang}, \citenamefont {Li}, \citenamefont {Wang}, \citenamefont
  {Ye},\ and\ \citenamefont {Jiang}}]{Lu:2012it}%
  \BibitemOpen
  \bibfield  {author} {\bibinfo {author} {\bibfnamefont {W.-T.}\ \bibnamefont
  {Lu}}, \bibinfo {author} {\bibfnamefont {S.-J.}\ \bibnamefont {Wang}},
  \bibinfo {author} {\bibfnamefont {W.}~\bibnamefont {Li}}, \bibinfo {author}
  {\bibfnamefont {Y.-L.}\ \bibnamefont {Wang}}, \bibinfo {author}
  {\bibfnamefont {C.-Z.}\ \bibnamefont {Ye}}, \ and\ \bibinfo {author}
  {\bibfnamefont {H.}~\bibnamefont {Jiang}},\ }\bibfield  {title} {\enquote
  {\bibinfo {title} {{Fano-type resonance through a time-periodic potential in
  graphene}},}\ }\href@noop {} {\bibfield  {journal} {\bibinfo  {journal}
  {Journal Of Applied Physics}\ }\textbf {\bibinfo {volume} {111}},\ \bibinfo
  {pages} {103717} (\bibinfo {year} {2012})}\BibitemShut {NoStop}%
\bibitem [{\citenamefont {Szab{\'o}}\ \emph {et~al.}(2013)\citenamefont
  {Szab{\'o}}, \citenamefont {Benedict}, \citenamefont {Czirj{\'a}k},\ and\
  \citenamefont {F{\"o}ldi}}]{Szabo:2013bn}%
  \BibitemOpen
  \bibfield  {author} {\bibinfo {author} {\bibfnamefont {L.~Z.}\ \bibnamefont
  {Szab{\'o}}}, \bibinfo {author} {\bibfnamefont {M.~G.}\ \bibnamefont
  {Benedict}}, \bibinfo {author} {\bibfnamefont {A.}~\bibnamefont
  {Czirj{\'a}k}}, \ and\ \bibinfo {author} {\bibfnamefont {P.}~\bibnamefont
  {F{\"o}ldi}},\ }\bibfield  {title} {\enquote {\bibinfo {title} {{Relativistic
  electron transport through an oscillating barrier: Wave-packet generation and
  Fano-type resonances}},}\ }\href@noop {} {\bibfield  {journal} {\bibinfo
  {journal} {Physical Review B}\ }\textbf {\bibinfo {volume} {88}},\ \bibinfo
  {pages} {075438} (\bibinfo {year} {2013})}\BibitemShut {NoStop}%
\bibitem [{\citenamefont {Zhu}\ \emph {et~al.}(2015)\citenamefont {Zhu},
  \citenamefont {Dai},\ and\ \citenamefont {Guo}}]{Zhu:2015bd}%
  \BibitemOpen
  \bibfield  {author} {\bibinfo {author} {\bibfnamefont {R.}~\bibnamefont
  {Zhu}}, \bibinfo {author} {\bibfnamefont {J.-H.}\ \bibnamefont {Dai}}, \ and\
  \bibinfo {author} {\bibfnamefont {Y.}~\bibnamefont {Guo}},\ }\bibfield
  {title} {\enquote {\bibinfo {title} {{Fano resonance in the nonadiabatically
  pumped shot noise of a time-dependent quantum well in a two-dimensional
  electron gas and graphene}},}\ }\href@noop {} {\bibfield  {journal} {\bibinfo
   {journal} {Journal Of Applied Physics}\ }\textbf {\bibinfo {volume} {117}},\
  \bibinfo {pages} {164306} (\bibinfo {year} {2015})}\BibitemShut {NoStop}%
\bibitem [{\citenamefont {Korniyenko}\ \emph
  {et~al.}(2016{\natexlab{a}})\citenamefont {Korniyenko}, \citenamefont
  {Shevtsov},\ and\ \citenamefont {L{\"o}fwander}}]{Korniyenko:2016ct}%
  \BibitemOpen
  \bibfield  {author} {\bibinfo {author} {\bibfnamefont {Y.}~\bibnamefont
  {Korniyenko}}, \bibinfo {author} {\bibfnamefont {O.}~\bibnamefont
  {Shevtsov}}, \ and\ \bibinfo {author} {\bibfnamefont {T.}~\bibnamefont
  {L{\"o}fwander}},\ }\bibfield  {title} {\enquote {\bibinfo {title} {{Resonant
  second-harmonic generation in a ballistic graphene transistor with an
  ac-driven gate}},}\ }\href@noop {} {\bibfield  {journal} {\bibinfo  {journal}
  {Physical Review B}\ }\textbf {\bibinfo {volume} {93}},\ \bibinfo {pages}
  {035435} (\bibinfo {year} {2016}{\natexlab{a}})}\BibitemShut {NoStop}%
\bibitem [{\citenamefont {Korniyenko}\ \emph
  {et~al.}(2016{\natexlab{b}})\citenamefont {Korniyenko}, \citenamefont
  {Shevtsov},\ and\ \citenamefont {L{\"o}fwander}}]{Korniyenko:2016hg}%
  \BibitemOpen
  \bibfield  {author} {\bibinfo {author} {\bibfnamefont {Y.}~\bibnamefont
  {Korniyenko}}, \bibinfo {author} {\bibfnamefont {O.}~\bibnamefont
  {Shevtsov}}, \ and\ \bibinfo {author} {\bibfnamefont {T.}~\bibnamefont
  {L{\"o}fwander}},\ }\bibfield  {title} {\enquote {\bibinfo {title}
  {{Nonlinear response of a ballistic graphene transistor with an ac-driven
  gate: High harmonic generation and terahertz detection}},}\ }\href@noop {}
  {\bibfield  {journal} {\bibinfo  {journal} {Physical Review B}\ }\textbf
  {\bibinfo {volume} {94}},\ \bibinfo {pages} {125445} (\bibinfo {year}
  {2016}{\natexlab{b}})}\BibitemShut {NoStop}%
\bibitem [{\citenamefont {BAGWELL}\ and\ \citenamefont
  {LAKE}(1992)}]{BAGWELL:1992wx}%
  \BibitemOpen
  \bibfield  {author} {\bibinfo {author} {\bibfnamefont {P.}~\bibnamefont
  {Bagwell}}\ and\ \bibinfo {author} {\bibfnamefont {R.}~\bibnamefont {Lake}},\
  }\bibfield  {title} {\enquote {\bibinfo {title} {{Resonances in Transmission
  through an Oscillating Barrier}},}\ }\href@noop {} {\bibfield  {journal}
  {\bibinfo  {journal} {Physical Review B}\ }\textbf {\bibinfo {volume} {46}},\
  \bibinfo {pages} {15329--15336} (\bibinfo {year} {1992})}\BibitemShut
  {NoStop}%
\bibitem [{\citenamefont {Pedersen}\ and\ \citenamefont
  {BUTTIKER}(1998)}]{1998PhRvB..5812993P}%
  \BibitemOpen
  \bibfield  {author} {\bibinfo {author} {\bibfnamefont {M.~H.}\ \bibnamefont
  {Pedersen}}\ and\ \bibinfo {author} {\bibfnamefont {M.}~\bibnamefont
  {B\"uttiker}},\ }\bibfield  {title} {\enquote {\bibinfo {title} {{Scattering
  theory of photon-assisted electron transport}},}\ }\href@noop {} {\bibfield
  {journal} {\bibinfo  {journal} {Physical Review B}\ }\textbf {\bibinfo
  {volume} {58}},\ \bibinfo {pages} {12993--13006} (\bibinfo {year}
  {1998})}\BibitemShut {NoStop}%
\bibitem [{\citenamefont {Platero}\ and\ \citenamefont
  {Aguado}(2004)}]{Platero:2004ep}%
  \BibitemOpen
  \bibfield  {author} {\bibinfo {author} {\bibfnamefont {G.}~\bibnamefont
  {Platero}}\ and\ \bibinfo {author} {\bibfnamefont {R.}~\bibnamefont
  {Aguado}},\ }\bibfield  {title} {\enquote {\bibinfo {title} {{Photon-assisted
  transport in semiconductor nanostructures}},}\ }\href@noop {} {\bibfield
  {journal} {\bibinfo  {journal} {Physics Reports-Review Section Of Physics
  Letters}\ }\textbf {\bibinfo {volume} {395}},\ \bibinfo {pages} {1--157}
  (\bibinfo {year} {2004})}\BibitemShut {NoStop}%
\bibitem [{\citenamefont {Kohler}\ \emph {et~al.}(2005)\citenamefont {Kohler},
  \citenamefont {Lehmann},\ and\ \citenamefont {Hanggi}}]{2005PhR...406..379K}%
  \BibitemOpen
  \bibfield  {author} {\bibinfo {author} {\bibfnamefont {S.}~\bibnamefont
  {Kohler}}, \bibinfo {author} {\bibfnamefont {J.}~\bibnamefont {Lehmann}}, \
  and\ \bibinfo {author} {\bibfnamefont {P.}~\bibnamefont {Hanggi}},\
  }\bibfield  {title} {\enquote {\bibinfo {title} {{Driven quantum transport on
  the nanoscale}},}\ }\href@noop {} {\bibfield  {journal} {\bibinfo  {journal}
  {Physics Reports-Review Section Of Physics Letters}\ }\textbf {\bibinfo
  {volume} {406}},\ \bibinfo {pages} {379--443} (\bibinfo {year}
  {2005})}\BibitemShut {NoStop}%
\bibitem [{\citenamefont {Katsnelson}(2006)}]{Katsnelson:2006gl}%
  \BibitemOpen
  \bibfield  {author} {\bibinfo {author} {\bibfnamefont {M.~I.}\ \bibnamefont
  {Katsnelson}},\ }\bibfield  {title} {\enquote {\bibinfo {title}
  {{Zitterbewegung, chirality, and minimal conductivity in graphene}},}\
  }\href@noop {} {\bibfield  {journal} {\bibinfo  {journal} {The European
  Physical Journal B}\ }\textbf {\bibinfo {volume} {51}},\ \bibinfo {pages}
  {157--160} (\bibinfo {year} {2006})}\BibitemShut {NoStop}%
\bibitem [{\citenamefont {Tworzyd{\l}o}\ \emph {et~al.}(2006)\citenamefont
  {Tworzyd{\l}o}, \citenamefont {Trauzettel}, \citenamefont {Titov},
  \citenamefont {Rycerz},\ and\ \citenamefont {Beenakker}}]{Tworzydio:2006hw}%
  \BibitemOpen
  \bibfield  {author} {\bibinfo {author} {\bibfnamefont {J.}~\bibnamefont
  {Tworzyd{\l}o}}, \bibinfo {author} {\bibfnamefont {B.}~\bibnamefont
  {Trauzettel}}, \bibinfo {author} {\bibfnamefont {M.}~\bibnamefont {Titov}},
  \bibinfo {author} {\bibfnamefont {A.}~\bibnamefont {Rycerz}}, \ and\ \bibinfo
  {author} {\bibfnamefont {C.}~\bibnamefont {Beenakker}},\ }\bibfield  {title}
  {\enquote {\bibinfo {title} {{Sub-Poissonian Shot Noise in Graphene}},}\
  }\href@noop {} {\bibfield  {journal} {\bibinfo  {journal} {Physical Review
  Letters}\ }\textbf {\bibinfo {volume} {96}},\ \bibinfo {pages} {246802}
  (\bibinfo {year} {2006})}\BibitemShut {NoStop}%
\bibitem [{\citenamefont {Danneau}\ \emph {et~al.}(2008)\citenamefont
  {Danneau}, \citenamefont {Wu}, \citenamefont {Craciun}, \citenamefont
  {Russo}, \citenamefont {Tomi}, \citenamefont {Salmilehto}, \citenamefont
  {Morpurgo},\ and\ \citenamefont {Hakonen}}]{Danneau:2008kg}%
  \BibitemOpen
  \bibfield  {author} {\bibinfo {author} {\bibfnamefont {R.}~\bibnamefont
  {Danneau}}, \bibinfo {author} {\bibfnamefont {F.}~\bibnamefont {Wu}},
  \bibinfo {author} {\bibfnamefont {M.~F.}\ \bibnamefont {Craciun}}, \bibinfo
  {author} {\bibfnamefont {S.}~\bibnamefont {Russo}}, \bibinfo {author}
  {\bibfnamefont {M.~Y.}\ \bibnamefont {Tomi}}, \bibinfo {author}
  {\bibfnamefont {J.}~\bibnamefont {Salmilehto}}, \bibinfo {author}
  {\bibfnamefont {A.~F.}\ \bibnamefont {Morpurgo}}, \ and\ \bibinfo {author}
  {\bibfnamefont {P.~J.}\ \bibnamefont {Hakonen}},\ }\bibfield  {title}
  {\enquote {\bibinfo {title} {{Shot Noise in Ballistic Graphene}},}\
  }\href@noop {} {\bibfield  {journal} {\bibinfo  {journal} {Phys Rev Lett}\
  }\textbf {\bibinfo {volume} {100}},\ \bibinfo {pages} {196802} (\bibinfo
  {year} {2008})}\BibitemShut {NoStop}%
\bibitem [{\citenamefont {Blanter}\ and\ \citenamefont
  {BUTTIKER}(2000)}]{2000PhR...336....1B}%
  \BibitemOpen
  \bibfield  {author} {\bibinfo {author} {\bibfnamefont {Y.~M.}\ \bibnamefont
  {Blanter}}\ and\ \bibinfo {author} {\bibfnamefont {M.}~\bibnamefont
  {B\"uttiker}},\ }\bibfield  {title} {\enquote {\bibinfo {title} {{Shot noise in
  mesoscopic conductors}},}\ }\href@noop {} {\bibfield  {journal} {\bibinfo
  {journal} {Physics Reports}\ }\textbf {\bibinfo {volume} {336}},\ \bibinfo
  {pages} {1--166} (\bibinfo {year} {2000})}\BibitemShut {NoStop}%
\bibitem [{\citenamefont {Parmentier}\ \emph {et~al.}(2016)\citenamefont
  {Parmentier}, \citenamefont {Serkovic-Loli}, \citenamefont {Roulleau},\ and\
  \citenamefont {Glattli}}]{Parmentier:2016ec}%
  \BibitemOpen
  \bibfield  {author} {\bibinfo {author} {\bibfnamefont {F.~D.}\ \bibnamefont
  {Parmentier}}, \bibinfo {author} {\bibfnamefont {L.~N.}\ \bibnamefont
  {Serkovic-Loli}}, \bibinfo {author} {\bibfnamefont {P.}~\bibnamefont
  {Roulleau}}, \ and\ \bibinfo {author} {\bibfnamefont {D.~C.}\ \bibnamefont
  {Glattli}},\ }\bibfield  {title} {\enquote {\bibinfo {title}
  {{Photon-Assisted Shot Noise in Graphene in the Terahertz Range}},}\
  }\href@noop {} {\bibfield  {journal} {\bibinfo  {journal} {Physical Review
  Letters}\ }\textbf {\bibinfo {volume} {116}},\ \bibinfo {pages} {227401}
  (\bibinfo {year} {2016})}\BibitemShut {NoStop}%
\bibitem [{\citenamefont {Laitinen}\ \emph {et~al.}(2016)\citenamefont
  {Laitinen}, \citenamefont {Paraoanu}, \citenamefont {Oksanen}, \citenamefont
  {Craciun}, \citenamefont {Russo}, \citenamefont {Sonin},\ and\ \citenamefont
  {Hakonen}}]{Laitinen:2016ht}%
  \BibitemOpen
  \bibfield  {author} {\bibinfo {author} {\bibfnamefont {A.}~\bibnamefont
  {Laitinen}}, \bibinfo {author} {\bibfnamefont {G.~S.}\ \bibnamefont
  {Paraoanu}}, \bibinfo {author} {\bibfnamefont {M.}~\bibnamefont {Oksanen}},
  \bibinfo {author} {\bibfnamefont {M.~F.}\ \bibnamefont {Craciun}}, \bibinfo
  {author} {\bibfnamefont {S.}~\bibnamefont {Russo}}, \bibinfo {author}
  {\bibfnamefont {E.}~\bibnamefont {Sonin}}, \ and\ \bibinfo {author}
  {\bibfnamefont {P.}~\bibnamefont {Hakonen}},\ }\bibfield  {title} {\enquote
  {\bibinfo {title} {{Contact doping, Klein tunneling, and asymmetry of shot
  noise in suspended graphene}},}\ }\href@noop {} {\bibfield  {journal}
  {\bibinfo  {journal} {Physical Review B}\ }\textbf {\bibinfo {volume} {93}},\
  \bibinfo {pages} {115413} (\bibinfo {year} {2016})}\BibitemShut {NoStop}%
\bibitem [{\citenamefont {Hammer}\ and\ \citenamefont
  {Belzig}(2013)}]{Hammer:2013gb}%
  \BibitemOpen
  \bibfield  {author} {\bibinfo {author} {\bibfnamefont {J.}~\bibnamefont
  {Hammer}}\ and\ \bibinfo {author} {\bibfnamefont {W.}~\bibnamefont
  {Belzig}},\ }\bibfield  {title} {\enquote {\bibinfo {title} {{Scattering
  approach to frequency-dependent current noise in Fabry-P{\'e}rot graphene
  devices}},}\ }\href@noop {} {\bibfield  {journal} {\bibinfo  {journal}
  {Physical Review B}\ }\textbf {\bibinfo {volume} {87}},\ \bibinfo {pages}
  {125422} (\bibinfo {year} {2013})}\BibitemShut {NoStop}%
\bibitem [{\citenamefont {Huard}\ \emph {et~al.}(2008)\citenamefont {Huard},
  \citenamefont {Stander}, \citenamefont {Sulpizio},\ and\ \citenamefont
  {Goldhaber-Gordon}}]{Huard:2008hx}%
  \BibitemOpen
  \bibfield  {author} {\bibinfo {author} {\bibfnamefont {B.}~\bibnamefont
  {Huard}}, \bibinfo {author} {\bibfnamefont {N.}~\bibnamefont {Stander}},
  \bibinfo {author} {\bibfnamefont {J.~A.}\ \bibnamefont {Sulpizio}}, \ and\
  \bibinfo {author} {\bibfnamefont {D.}~\bibnamefont {Goldhaber-Gordon}},\
  }\bibfield  {title} {\enquote {\bibinfo {title} {{Evidence of the role of
  contacts on the observed electron-hole asymmetry in graphene}},}\ }\href@noop
  {} {\bibfield  {journal} {\bibinfo  {journal} {Physical Review B}\ }\textbf
  {\bibinfo {volume} {78}},\ \bibinfo {pages} {121402} (\bibinfo {year}
  {2008})}\BibitemShut {NoStop}%
\bibitem [{\citenamefont {Beenakker}\ and\ \citenamefont
  {BUTTIKER}(1992)}]{1992PhRvB..46.1889B}%
  \BibitemOpen
  \bibfield  {author} {\bibinfo {author} {\bibfnamefont {C.}~\bibnamefont
  {Beenakker}}\ and\ \bibinfo {author} {\bibfnamefont {M.}~\bibnamefont
  {B\"uttiker}},\ }\bibfield  {title} {\enquote {\bibinfo {title} {{Suppression
  of shot noise in metallic diffusive conductors.}}}\ }\href@noop {} {\bibfield
   {journal} {\bibinfo  {journal} {Physical review. B, Condensed matter}\
  }\textbf {\bibinfo {volume} {46}},\ \bibinfo {pages} {1889--1892} (\bibinfo
  {year} {1992})}\BibitemShut {NoStop}%
\bibitem [{\citenamefont {Moskalets}\ and\ \citenamefont
  {BUTTIKER}(2004)}]{Moskalets:2004ct}%
  \BibitemOpen
  \bibfield  {author} {\bibinfo {author} {\bibfnamefont {M.}~\bibnamefont
  {Moskalets}}\ and\ \bibinfo {author} {\bibfnamefont {M.}~\bibnamefont
  {B\"uttiker}},\ }\bibfield  {title} {\enquote {\bibinfo {title} {{Floquet
  scattering theory for current and heat noise in large amplitude adiabatic
  pumps}},}\ }\href@noop {} {\bibfield  {journal} {\bibinfo  {journal}
  {Physical Review B}\ }\textbf {\bibinfo {volume} {70}},\ \bibinfo {pages}
  {245305} (\bibinfo {year} {2004})}\BibitemShut {NoStop}%
\end{thebibliography}

%

\end{document}